\documentstyle[11pt]{article}
\input epsf
\def\be{\begin{equation}}
\def\ee{\end{equation}}
\def\bea{\begin{eqnarray}}
\def\eea{\end{eqnarray}}
\begin{document}
\begin{titlepage}
\begin{center}
\baselineskip24pt
{\Large\bf
The importance of initial-final state correlations 
for the formation of fragments in heavy ion collisions} 
\vskip 1 cm
P.-B. Gossiaux$^{1,2}$ and J. Aichelin$^1$\\ 
{\it$^1$
SUBATECH\\ Universit\'e de Nantes, EMN, IN2P3/CNRS
F-44072 Nantes, France}
\vskip .5 cm
{\it$^2$
National Superconducting Cyclotron Laboratory, MSU, USA}
\end{center}

\vskip 1 cm
\begin{abstract}
Using quantum molecular dynamics simulations, we investigate 
the formation of fragments in symmetric reactions between 
beam energies of E~=~30A~MeV and 600A~MeV. After a comparison
with existing data we investigate some observables relevant to
tackle equilibration: $\mbox{d}\sigma/\mbox{d}E_{rat}$, 
the double differential cross section 
$\mbox{d}^2\sigma/p_t\mbox{d}p_z\mbox{d}p_t$,... Apart maybe from very 
energetic ($\mbox{E}\ge\mbox{400A~MeV}$) and very central reactions, none 
of our simulations gives evidence that the system passes through a state of 
equilibrium. Later, we address the production mechanisms and find that, 
whatever the energy, nucleons finally entrained in a fragment exhibit 
strong initial-final state correlations, in coordinate as well as 
in momentum space. At high energy those correlations resemble the 
ones obtained in the participant-spectator model. At low energy the 
correlations are equally strong, but more complicated; they are a 
consequence of the Pauli blocking of the nucleon-nucleon collisions, 
the geometry, 
and the excitation energy. Studying a second set of time-dependent 
variables (radii, densities,...), we investigate in details how those 
correlations survive the reaction especially in central reactions where 
the nucleons have to pass through the whole system. 
It appears that some fragments are made of nucleons which were initially 
correlated, whereas others are formed by nucleons scattered during the 
reaction into the vicinity of a group of previously correlated nucleons.
\end{abstract}

\end{titlepage}
\newpage

\section{Introduction}
The multifragmentation of nuclei excited in collisions with protons or 
heavy ions is one of the most interesting and challenging topics in 
present-day heavy-ion physics. The production of intermediate mass 
fragments (IMF), which we define as objects with $3 \le Z \le 25$,
in collisions of a proton with heavy targets was first observed
about 40 years ago \cite{fried}. Using radiochemical methods, however,
a total cross section for fragmentation could not be determined and 
this process had been considered as quite rare and exotic. 

A decade ago
two findings, one experimental and one theoretical, have placed 
multifragmentation into the spotlight. Bombarding large target nuclei with 
heavy ions
Warwick et al. \cite{WW} found that many IMFs are produced simultaneously and
that multifragmentation is the dominant reaction channel at beam energies
larger than 35A~MeV. Observing that the mass yield curve approximately obeys
a power law $\sigma(A) \propto A^{-\tau}$, the Purdue group
\cite{hi} conjectured
that multifragmentation is a clear signature \cite{fi} for the phase transition
between a gaseous and a liquid phase of nuclear matter. 
This transition is predicted to occur around a density of $0.4\rho_0$,
$\rho_0$ being the normal nuclear-matter density.           

Since then, the study of multifragmentation has been considered
of such interest that special ($4\pi$) detectors
have been designed to inspect this process in detail. Today it
is a major research project at all heavy ion
accelerators. However, despite these extensive experimental efforts, 
the underlying physical mechanism remains unclear and the debate
is still quite controversial. This is due to the fact that the 
experiments have revealed many puzzling aspects. Some results are 
in perfect agreement with
the conjecture that the fragments are emitted from a globally thermalized
system. Others, however, can hardly be reconciled with this conjecture.
We will summarize these results shortly.

It has been observed that independently of the mass of the projectile-target 
combination, the {\it form} of the mass yield curve \cite{mor} is
almost identical from 30A~MeV up to the highest energies. The form observed
is that expected for formation of fragments in a liquid-gas phase
transition. Other observables show, however, that the underlying
process changes. At low beam energies the slopes of the
energy spectra of protons and IMFs agree, whereas, starting from about
100A~MeV, an increasing difference of the slopes is observed \cite{mi}.
This has been interpreted as an evidence that,
at low energy, the whole nucleus takes part in the multifragmentation
to the limit that the process can be described as a compound nucleus decay, 
whereas, at high energies, only the spectators' part contributes 
significantly \cite{mor}.

The backward energy spectra of fragments are found to be rather independent 
of the beam energy \cite{WW,WT} and of the projectile mass \cite{WW} 
(for $A_{\rm P} \le 20$). At each angle the spectra have the form 
expected for a thermal emission from a moving source. Thus single arm
experiments seem to indicate an emission from a thermal source.
If one compares the spectra observed under different angles 
\cite{WW,mi} one finds, however, that
the source velocity, the Coulomb barrier and the temperature depend
on the observation angle \cite{AH}. 
This contradicts the thermal source assumption. 

Recent experiments performed with a $4\pi$ detector have
revealed that the energy balance in central collisions is 
dominated by a large undirected flow component ($\beta
\approx 0.1 c$) and that the Coulomb repulsion and the possible
thermal energy represent only a minor part \cite{je}. Hence the
nucleus seems to explode.

The models advanced to describe 
multifragmentation invoke practically all imaginable physical processes.  
For a recent review we refer to \cite{mor}. Most of the 
phenomenological models only aim to explain/reproduce a few observables. 
In the last years, they have been superseded by 
numerically involved models which predict results for a multitude of 
observables simultaneously. Presently, these models can be
classified into two main categories.

First, statistical models \cite{gro,bon,sto} which neglect the
dynamics of the reaction by supposing that the system or a subsystem
reaches equilibrium in the course of the reaction
and maintains this equilibrium until its density has decayed
to approximately 0.4 $\rho_0$. According to this assumption,
all possible exit channels 
(consisting of nucleons and---possibly excited---fragments) 
are populated with equal probabilities. 
These models have revealed themselves to reproduce mass yields, fragment
multiplicities and fragment-mass correlations very accurately for 800 A~MeV
reactions \cite{bot} but fail to reproduce dynamical 
variables \cite{je,hs}, a fact not yet understood. 

A second type of approach is the {\em quantum molecular dynamics} 
(QMD) \cite{aic} model which follows the time-evolution of the
full multi-nucleon phase space distribution 
from the initial collision of projectile and target up
to the final formation of fragments. 
This model is more ambitious than the statistical approach. 
First, the QMD model permits to check whether complete equilibrium is 
reached for a given reaction. 
When thermal equilibrium is established, it
has been shown that the QMD model predicts the same fragment multiplicities
as the statistical models \cite{bbl}. However, it also {\it predicts} 
the density and the temperature which are just input parameters in  
thermal models. When complete equilibrium is not reached, it allows to 
investigate the reaction as well. For this reason, we 
claim that the QMD model goes beyond the thermal ones.
On the other hand, there are some drawbacks attached to 
the QMD approach. For instance, it usually requires more 
approximations than the statistical models, as discussed in 
section~\ref{qmd-model-description}... Nevertheless, let us mention from 
the beginning that QMD results reproduce the experimental data fairly well, 
as it will be detailed in section~\ref{QMD-valid}. 

Even though this agreement is primordial, 
the numerous physical processes included in the QMD code and 
the abundance of degrees of freedom very often prevent us to get a 
clear view of the relevant underlying physics and of the causal links.
But the physics of heavy ion collisions should not reduce to a black box 
with tuning parameters: At some point, one has to understand the reaction 
mechanisms in a coherent, comprehensive and preferably conceptual manner. 

It is precisely the goal of this paper to make some
steps towards a more global understanding of nucleus-nucleus 
collisions and their energy dependence (from basically 50A~MeV to 
600A~MeV), by analyzing the reaction as a function of   
time and by probing some specific time-correlations (that we shall 
sometimes simply denote ``correlations'' when no ambiguity exists). 
Quite generally, we shall speak of time-correlation if 
nucleons belonging to a specific class of emitted fragments  
have a different mean history than all nucleons considered 
indiscriminatingly.
A very good (but trivial) example of such a time-correlation can 
be found in the participant spectator (PS) model:
At high energy and finite impact parameter, experimental and 
theoretical results suggest that the system can be divided into 
two subsystems, the spectators and the participants. According to this 
model, the nucleons which compose the heavy fragments would 
mainly emerge from the spectator zones, so avoiding the fireball. 

One easily realizes that the analysis of time-correlations is
particularly useful to clarify whether the system has passed through a 
state of {\em global} equilibrium before (multi)fragmenting.
If it were the case, all the memory would be completely lost and  
time-correlations would disappear. However, we will find many of them,
strong and not only of geometrical nature. This indicates that 
inside the QMD approach, the system does not pass through a state of 
global equilibrium. 

Not only time-correlations act as a touchstone of the system's equilibration 
but they also play a first role in understanding (a)the mechanisms of 
fragment formation as well as (b)the way the memory of the entrance channel 
partly survives the high density/temperature phase of the heavy ion reaction.
We will illustrate this assertion by analyzing the time-evolution of 
some well-chosen quantities, like the radius of pre\footnote{See 
section~\ref{survey-section} for definition.}fragments, their internal
kinetic energy,...

The paper is organized as follows: In section~\ref{qmd-model-section},
we briefly introduce the QMD model (\ref{qmd-model-description}), 
discuss more extensively some of its aspects relevant in the 
context of fragmentation (\ref{qmd-model-mult}) and state
our conventions~\ref{convention-section}. Next (\ref{QMD-valid}), we 
validate the model by comparing it to recent 
experimental results obtained at MSU, GANIL and GSI.
In section~\ref{survey-section}, we present a survey of some typical 
reactions, what will help the reader to get a general idea of
the propagation of prefragments at intermediate times. In
Section~\ref{whole-therm-sect}, we investigate to what degree
the system achieves global equilibration. We present results for the 
stopping, the double differential cross section 
${\mbox{d}^2\sigma/p_t \mbox{d}p_t\mbox{d}p_z}$ and the mixing of
projectile and target nucleons, which suggest a small degree of equilibration.
In Section~\ref{init-fin-corr-section}, we detail the initial-final state 
correlations for symmetric systems between 50A~MeV and 400A~MeV.  
After outlining the method employed to construct these correlations
(\ref{detail-time-corr}), we address both cases of coordinate and
momentum spaces. We find strong correlations between the final fate of the 
nucleons and their initial position in phase space. 
Moreover, we have distinguished two classes of fragments (those finally 
located (a) around mid-rapidity and (b) around target or beam rapidity)... 
and discovered additional correlations.
In section~\ref{time-evol-of-reac}, we inspect how these correlations can 
be preserved during the time-evolution of the system, especially for the 
nucleons which traverse the whole reaction partner. We conclude 
our work in section~\ref{conclusions}.

\section{The QMD model}
\label{qmd-model-section}
\subsection{Description of the approach}
\label{qmd-model-description}
In the QMD model, the time evolution of the system is calculated by means 
of a generalized variational principle: After having chosen a test wave 
function $\phi$ (which contains time-dependent parameters), one evaluates
the action \be
S = \int\limits ^{t_2} _{t_1} {\cal L} [\phi, \phi^\ast] dt \ee
with the Lagrange functional ${\cal L}$ 
\begin{equation}
{\cal L} =\left\langle\phi \left\vert i\hbar {\frac{d }{dt}} - H \right\vert
\phi\right\rangle\;,
\end{equation}
where the total time derivative includes the derivation with respect to the
parameters. The time evolution of the parameters is then obtained by
requiring the action to be stationary under the allowed variation of the
wave function. This requirement leads to an Euler-Lagrange equation for each
time-dependent parameter.

If the true solution of the Schr\"odinger equation is contained in the
set of wave functions $\phi$, this
variational method of the action provides the {\em exact} solution of the 
Schr\"odinger equation. If the parameter space is too restricted, one obtains
the {\em closest} wave function to the exact solution (in that restricted 
parameter space).

The basic assumption of the QMD model is that a test wave function
of the form 
\begin{equation} 
\phi = \prod_{\alpha = 1}^{A_{P}+A_{T}} \phi_\alpha
\end{equation}
with 
\be
\phi_\alpha (x_1,t) = \left({\frac{2 }{L\pi}}\right)^{3/4}\, e^{-(x_1 -
x_\alpha(t) -p_\alpha(t) t/m)^2/L} \,e^{i(x_1-x_\alpha(t)) p_\alpha(t)}\,
e^{-i p_\alpha^2(t)t/2m} \ee
and
\be
\hbar,c =1 
\ee
is a good approximation to the nuclear wave function. 
Neglecting antisymmetrization is the most drastic approximation of the model
\footnote{There are two attempts to go beyond the standard QMD approach 
by constructing an antisymmetrized molecular dynamics model
\cite{fm,ho}. These approaches are not mature yet.
They gave a lot of insight into time evolution
of small systems at very low energies but they cannot be 
used for comparison with the most interesting fragmentation experiments 
due to two yet unsolved problems:
(1) The Slater determinant has $N!$ terms. This presently limits 
---despite several approximations---calculations to systems 
with 80 nucleons at the very most. (2) A consistent way to treat the
collision term (the imaginary part of Br\"uckner g-matrix), which is 
crucial for the time evolution of reactions at higher energies, 
has not been found yet. Whether the use of Jastrow correlated 
wave functions considered in \cite{fmm} instead of Slater determinants 
will improve the situation has to be seen.}, as,
for instance, all properties related to shell structures cannot 
be accounted for. 

The wave function $\phi_{\alpha}$ has two time-dependent parameters: 
$x_\alpha(t), p_\alpha(t)$, while L is fixed. The initial values of 
these parameters are chosen in such a way that the ensemble of $A_T$ + $A_P$
nucleons gives proper densities and momentum distributions of the projectile 
and target nuclei. 

For the coherent states $\phi_\alpha$ and the Hamiltonian 
\be
H = \sum_i T_i + {\frac{1}{2}} \sum_{ij} V_{ij}\;,
\ee
where $T_i$ is the kinetic energy and $V_{ij}$ the potential
energy between two nucleons, the Lagrangian and the variation can be 
easily evaluated: 
\be
{\cal L} = \sum_\alpha \biggl[\dot{x_\alpha} p_\alpha - {1\over 2}\sum _{\beta} \langle
V_{\alpha\beta}\rangle - {\frac{3 }{2Lm}} \biggr]\ee
\be
\dot{\bar{x}}_{\alpha} = {\frac{p_\alpha }{m}} + \nabla_{p_\alpha} \sum_\beta
\langle V_{\alpha\beta}\rangle \ee
\be
\dot{p}_{\alpha} = - \nabla_{\bar x_\alpha} \sum _\beta \langle
V_{\alpha\beta}\rangle \ee
with 
\be
\displaystyle {\bar x_\alpha} = x_\alpha + {\frac{p_\alpha }{m}} t
\ee
and 
\be 
<V_{\alpha\beta}> = \int d^3x_1d^3x_2<\phi_\alpha\phi_\beta\vert
V(x_1,x_2)\vert\phi_\alpha\phi_\beta>.
\ee

These are the time evolution equations to be solved numerically. 
The interaction $V_{ij}$ is taken as the real part of the Br\"uckner g-matrix
supplemented by the Coulomb interaction.
If the energy is sufficiently high, the g-matrix becomes complex and the
imaginary part acts like a cross section. Details may be found in \cite{aic}. 

The variational approach reduces the complicated task to follow
the time evolution of a n-body wave function to the resolution of 
$6\,n$ coupled differential equations for the centroids of the coherent 
state wave
functions in coordinate and momentum space. In the following,
we shall sloppily call these centroids position and momentum of the 
nucleons. However, for any rigorous interpretation, one has to keep in 
mind that these are just parameters of a wave function
which indeed obeys the uncertainty principle. 

This approach also allows a very convenient definition of clusters: At the
end of the simulation the average phase space occupation is quite low.
Hence only nucleons with mutual interactions, i.e. the ones forming
a fragment are close together in coordinate space. 
A simple minimum-spanning-tree algorithm, applied to the
centroids of the coherent states, permits to define the clusters. 
If the average phase-space occupation is sufficiently low the
result is quite independent of the minimum-spanning-tree radius.

The QMD model has been 
extensively discussed in a recent Physics Reports~\cite{aic}. Later 
the results have been updated in other reviews \cite{aich,har}; 
therefore we do not repeat further details here. Let us just
discuss a few important points concerning fragmentation
and time-correlations.

\subsection{Aspects of the QMD model specific to fragmentation 
and time-correlations}
\label{qmd-model-mult}
1. We employ a direct product of coherent states. Hence, one may in 
principle find more than one particle in a unit cell $h^3$ 
of the phase space although this is forbidden for the one-body phase space
density by the Liouville theorem. To avoid a
possible overoccupation one has sometimes introduced a potential 
depending on the relative momenta and the relative positions of the
particles. This so called Pauli potential provides a well defined 
nuclear ground state which corresponds to the minimum of the 
Hamiltonian used in the calculation. For an isolated nucleus,
the nucleons carry their proper Fermi momentum but they do not
move relatively to the center of mass (${dx_i / dt} = 0$). 
Hence the price one has to pay is the crystallization of the cold nucleus:  
Nothing else but a collective motion is possible. Indeed, if just one nucleon
is moved, it enters an already occupied cell. Due to this crystallization
the fragments are excited for a much longer time than in the standard QMD
(the one we use in this paper), where the fragments are practically cold 
after 180~fm/c.
Therefore, a so called afterburner, i.e. a statistical evaporation program, 
has to be used at the end of the reaction to deexcite the
hot fragments into their ground state. 
These two approaches (QMD with Pauli potential + afterburner and standard 
QMD) are conceptually different and naturally lead to fragmentation patterns 
that also differ quite substantially \cite{bet}.

2. The QMD parameterization turns out to be very
useful if one is interested in following the flow of matter.
Indeed, inside the model, the one-body density is
\begin{equation}  
\rho\left(x,t\right)=\sum_{\alpha=1}^{A_{P}+A_{T}}
\left|\phi_{\alpha}\left(x,t\right)\right|^{2}\;.
\end{equation}
As all the $\phi_{\alpha}$ are peaked around their centroids, 
one easily figures out how nuclear matter evolves from one point to 
another by just following the centroid distribution.
For instance, tagging nucleons as projectile or target-like, one can 
study how efficiently they mix in the course
of the reaction. Of course, in nature, particles are indistinguishable 
and attaching such a flag of origin to each particle seems a bit
unrealistic. However, by going to the Wigner density formalism, we can 
interpret the time evolution of the Schr\"odinger equation as the flow of 
the phase space density. This allows us to investigate the (mean) origin 
of the matter contained in the final fragments. In this context, 
particles merely serve as an expedient to calculate that flow of matter
\footnote{One has however to keep in mind that the interpretation of the 
Wigner density as a phase space density can lead to some complications. 
For instance, the Wigner density is not positive definite.}.

3. Another delicate point where the unsatisfactory treatment of that 
indistinguishable character may cause
some trouble are the nucleon-nucleon collisions.
In standard QMD, the scattering angle in the nucleon-nucleon center of 
mass system ($\theta_{cm}$) is always taken lower than $\pi/2$.
At the large energy limit, this is justified, as we know that 
a transfer of large longitudinal momentum is {\em improbable}: In 
proton-neutron collisions, $\theta_{cm}$ almost never exceeds $\pi/2$. 
To assume a similar constrain in proton-proton and neutron-neutron collisions 
is therefore acceptable. In more technical terms, the nucleon-nucleon 
scattering amplitude basically contains 2 contributions: the t and the u 
channels. At high energy, the t channel dominates by far: the outgoing 
forward nucleon has a high probability to be the same particle as the 
ingoing forward nucleon. 

On the contrary, at lower energy, both scattering amplitudes 
associated with the t and u channels become independent of
$\theta_{cm}$, and the differential cross section is almost 
isotropic. Accordingly, one should in principle remove that 
$\theta_{cm}\le \pi/2$ constrain from the model and let 
$\theta_{cm}$ range from $0$ to $\pi$.
This modification has no consequence on the final state
of the reaction (number, masses and momenta of the fragments)
but may change dramatically the time-correlations and our
mental representation of the reaction evolution. 

Of course, one can also take the ``information'' point of view: a 
nucleon-nucleon collision happening exactly at 
$\theta_{cm}=\pi$ leaves the physical state unchanged\footnote{Physically, 
it even cannot be distinguished from a $\theta_{cm}=0$ collision.}. The 
forward (respectively backward) outgoing 
nucleon transmits the {\em memory} of the forward (respectively backward)
ingoing nucleon. Only a collision with a finite 
$\sin\left(\theta_{cm}\right)$ is able to change the physical state. Hence, 
to study how efficiently the collisions destroy the memory of the system, 
maintaining $\theta_{cm}\le \pi/2$ is not so unrealistic as it may appear 
at first sight. 

Which one of these two conventions should we adopt? 
At very high energy ($E\ge\mbox{1 GeV}$), this question is
irrelevant. On the other hand, at very low energy ($E\le\mbox{100 MeV}$)
one can invoke the Pauli-blocking of the collisions which should be strong 
enough to reduce the number of collisions significantly and solve the 
dilemma. Unfortunately, 
for energies ranging between those two values, the problem shows up.
We have chosen the following approach: 
for the main discussion, we shall stick to the standard QMD and 
$\theta_{cm}\in[0,\pi/2]$ (called hereafter ``model (A)'').
Occasionally, some crucial quantities will be reevaluated in the 
``$\theta_{cm}\in[0,\pi]$'' version of QMD (called hereafter ``model (B)'') 
and compared to the results obtained in model (A). 

Hereafter, we state the other underlying conventions and clarify 
the few non-standard terms used in this paper: 

\subsection{Conventions}
\label{convention-section}
\begin{itemize}
\item
the results are displayed in the nucleus nucleus center of mass system
\item
the beam axis is in z direction, the impact parameter points into the
x direction
\item
the projectile has initially a positive z momentum and is displaced
(for non zero impact parameters) in the positive x direction
\item
A soft equation of state \cite{aic} has been used. Consequently, 
for some observables (i.e. flow), one does not expect to have 
{\em quantitative} agreement with experiment.
\item
Whenever they appear, the following symbols are
to be understood as defined hereafter:
\begin{eqnarray*}
b_{<}&\equiv&\mbox{3 fm for all systems}\\
b_{>}&\equiv&\left\{\begin{array}{l}\mbox{6~fm for Xe~+~Sn}\\
\mbox{8~fm for Au~+~Au}\end{array}\right.\\
\mbox{LF}&\equiv&\mbox{light fragment: any single or fragment of mass 
$2\le A \le 4$}\\
\mbox{HF}&\equiv&\mbox{heavy fragment: any fragment of mass $A \ge 5$}
\end{eqnarray*}
\end{itemize}

\subsection{The QMD model compared to experiment}
\label{QMD-valid}
As we have explained, the quest for correlations between the initial and 
the final state of a heavy ion reaction is meaningful only
if the model employed contains the essential physics.
To check that QMD basically satisfies this requirement, extensive 
calculations have been performed and compared
\footnote{In fact, the comparison between experiment and
theory is only made possible by very extensive filter programs.
They determine which particles of a numerical simulation would have
been detected by the actual apparatus. For this decision, one has
to develop a detailed understanding of how the detector reacts to double 
hits, energy thresholds, etc. } to the most demanding data recently 
measured with 4 $\pi$ detectors at GSI, MSU and GANIL. The data
obtained at these facilities are to a large extend still preliminary,
so that the conclusions we draw should not be considered as definitive. 
 
\begin{figure}[t]
$$
\epsfbox{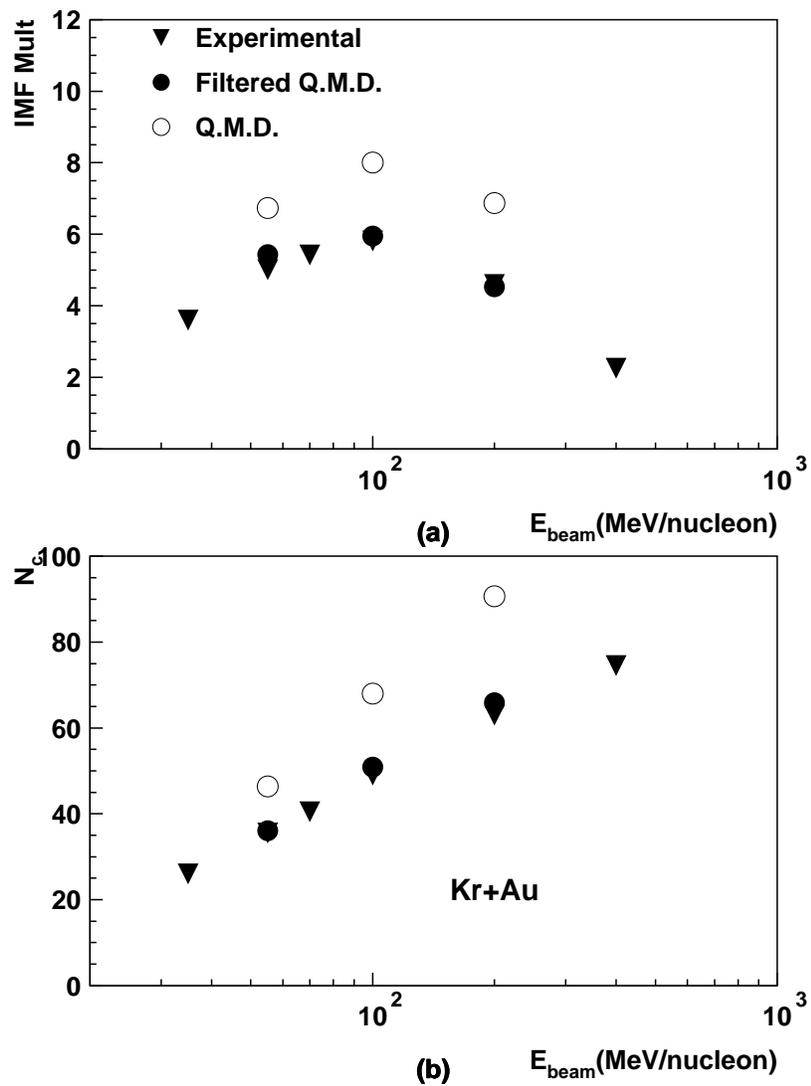}
$$
\caption{Multiplicity of intermediate mass fragments (IMF) (a)
and number of charged particles ($\mbox{N}_{C}$) (b) as a function of 
the beam energy for Kr + Au reactions.}
\label{imf_nc_msu_comp_part2}
\end{figure}

In figure~\ref{imf_nc_msu_comp_part2}, we present two quantities
fundamental for the multifragmentation: the intermediate mass 
fragment (IMF) multiplicity and the number of charged particles 
($\mbox{N}_{C}$) as a function of the beam energy (E) for the Kr~+~Au 
system at 3 different energies (from 55A~MeV up to 200A~MeV). In this 
figure, the experimental data obtained by the MSU collaboration \cite{MSU}
and the filtered QMD-calculations are confronted, with good overall 
agreement.

\begin{figure}[t]
$$
\epsfbox{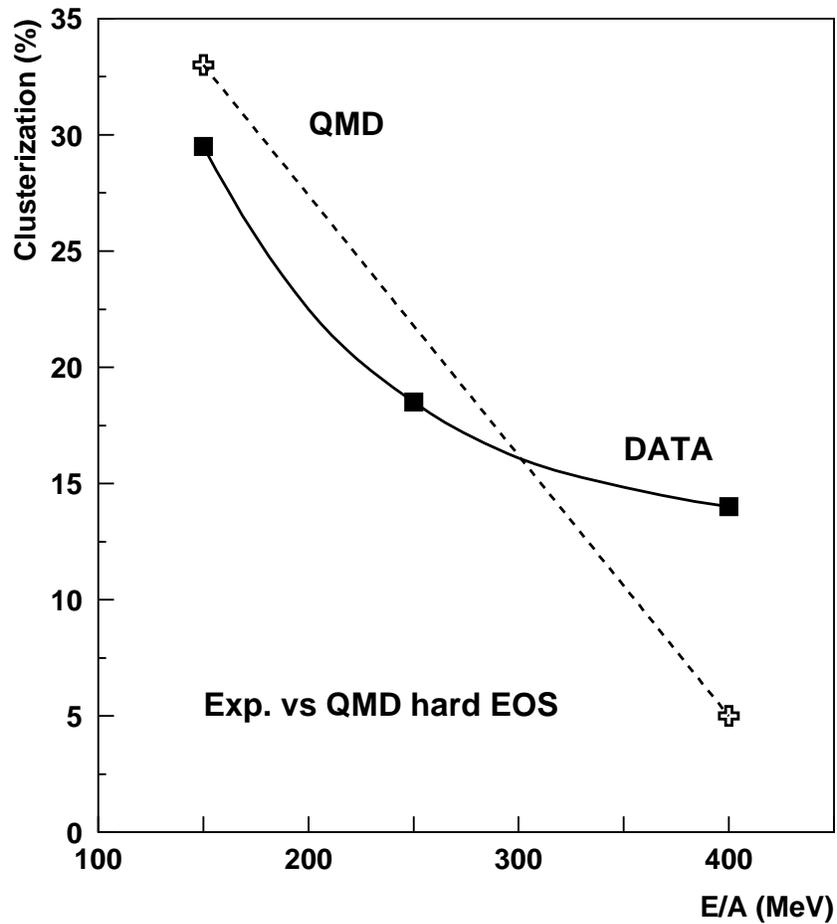}
$$
\caption{Percentage of nucleons bound in clusters ($\mbox{Z}\ge 3$) as a 
function of the beam energy as calculated with QMD and compared to the FOPI 
experiments by the FOPI collaboration.}
\label{zclust}
\end{figure}

In figure~\ref{zclust}, we present another important quantity: the 
percentage of nucleons bound in fragments $Z\ge 3$. It has been 
measured experimentally for the Au~+~Au system by the FOPI collaboration.
Although not displayed, the maximum percentage is obtained for beam energies 
around 50A~MeV. Also, one notices a rapid decrease at high energy.
QMD reproduces the low-energy value, but for 
E~=~400A~MeV, this is no longer the case... In fact, it is well-known
that for energies larger than 400A~MeV, the present version
of the QMD model fails to describe the IMF multiplicity \cite{bb} 
(although the dependence of the number of nucleons bound in clusters as a 
function of the $\mbox{N}_{C}$ is correctly reproduced). The reason is not 
yet understood... However, one could 
demonstrate that this failure is not due to the thermal properties of 
the QMD nuclei but more likely related to the low momentum transfer to the 
spectators \cite{bbl}. It is also true that statistical models 
succeed in describing these mass yields \cite{bot}. Nevertheless, they take 
the freedom to adjust the excitation energy of the system by hand,
while in the QMD approach, it is completely determined 
by the solution of the time-evolution equations. 

\begin{figure}[t]
$$
\epsfbox{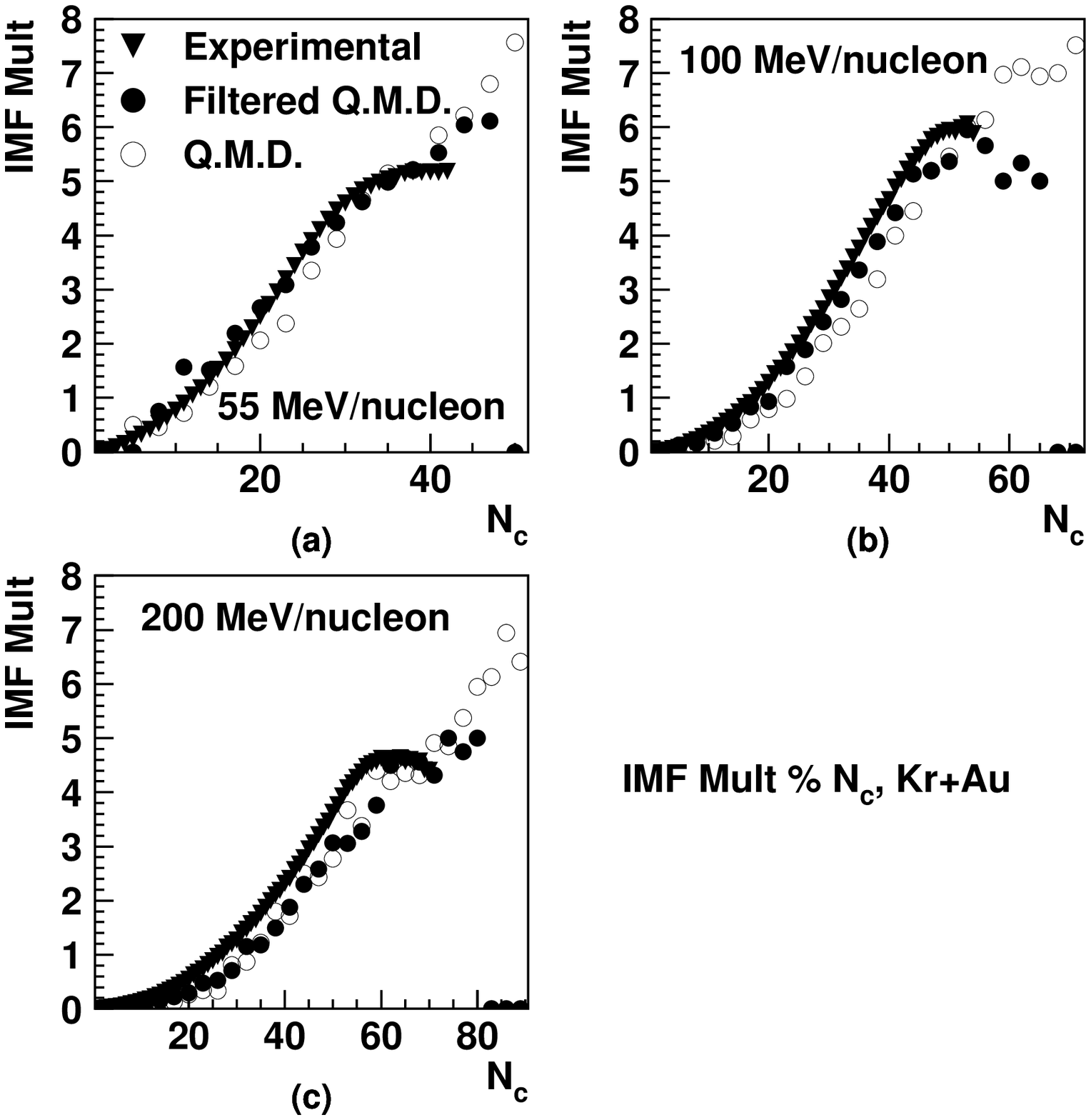}
$$
\caption{Average multiplicity of intermediate mass fragments (IMF) 
as a function of the number of charged particle ($\mbox{N}_C$) for 
Kr~+~Au reactions at three different beam energies 
(55, 100 and 200A~MeV). We compare the experimental values with the 
filtered QMD-results.}
\label{imf_nc_msu_comp_part1}
\end{figure}

\begin{figure}[t]
$$
\epsfbox{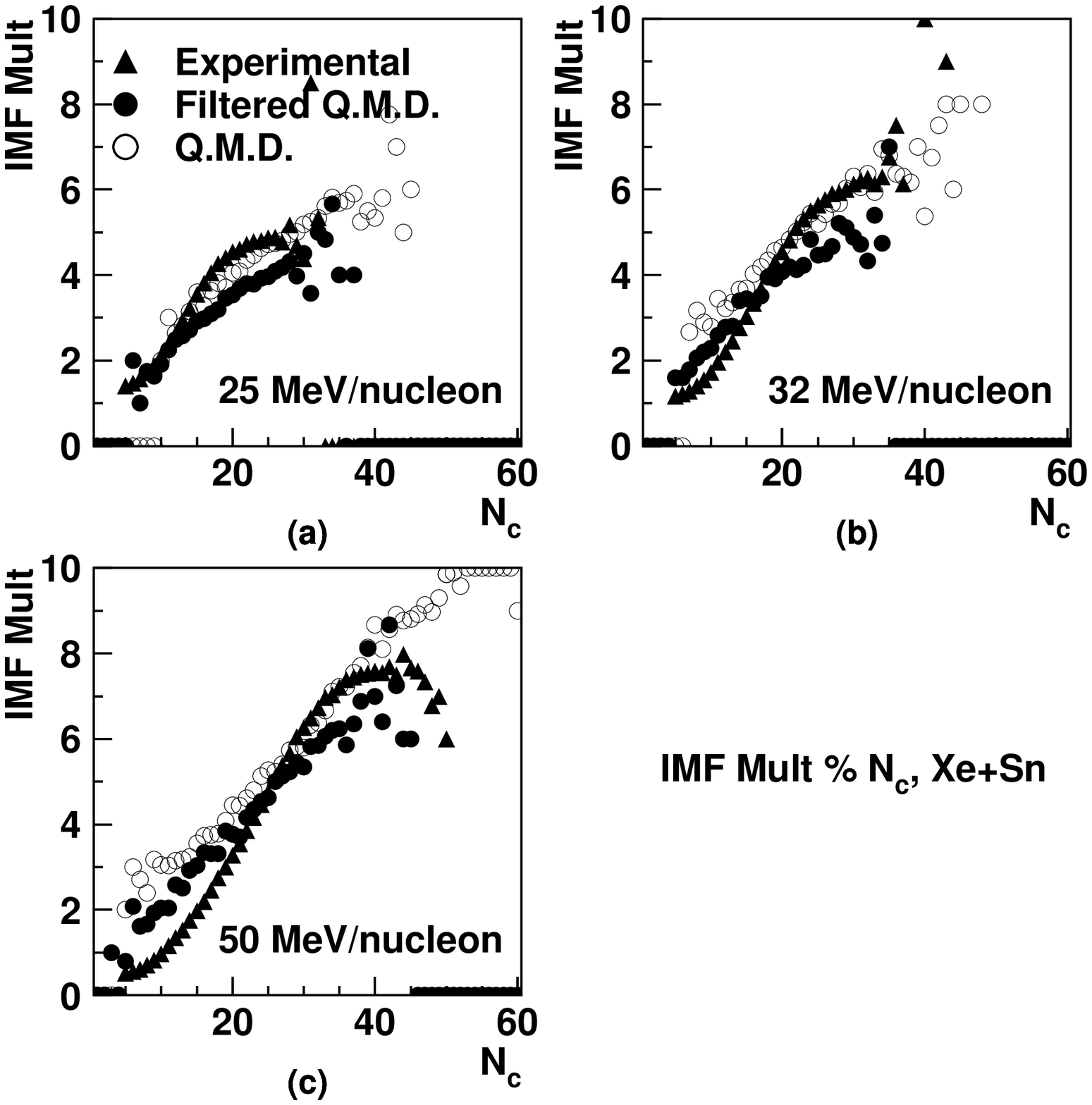}
$$
\caption{Average multiplicity of intermediate mass fragments (IMF)
as a function of the number of charged particle ($\mbox{N}_C$) for 
Xe~+~Sn reactions at three different beam energies (25, 32 and 50A~MeV). 
We compare the experimental values with the filtered QMD-results.}
\label{indra_lmfnc_val3e3.ps}
\end{figure}

In figure~\ref{imf_nc_msu_comp_part1}, we present a more refined signature 
of the multifragmentation: the IMF multiplicity as a function of 
$\mbox{N}_{C}$, for some of the reactions studied in  
figure~\ref{imf_nc_msu_comp_part2}. At all energies 
(from 55A~MeV up to 200A~MeV), a good agreement
between theory and experiment is achieved. 
A fair agreement has also been obtained for the
recent measurements performed with the INDRA detector at GANIL 
(figure~\ref{indra_lmfnc_val3e3.ps}). The deviations 
observed at small $\mbox{N}_{C}$ may have two origins: 
 (1) the filter routines are still preliminary and/or 
 (2) collisions at large impact parameter are more difficult
to describe theoretically. 

\begin{figure}[t]
$$
\epsfbox{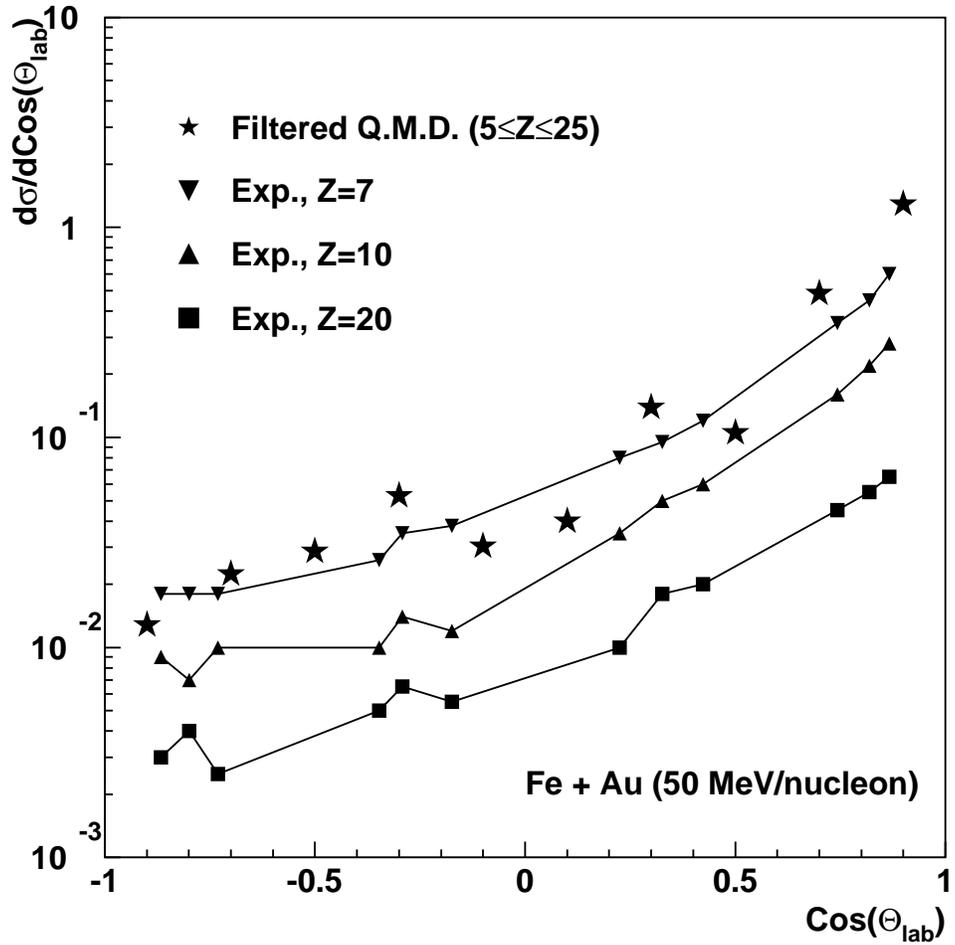}
$$
\caption{Angular distribution of fragments in the Fe~+~Au reactions 
at 50A~MeV. The experimental angular distribution for Z~=~7,~10 and 20
are compared with filtered QMD results for the angular distribution
of all fragments $5\le\mbox{Z}\le 25$.} 
\label{gsang}
\end{figure}

An even more detailed investigation of a 50A~MeV reaction is
available for the asymmetric system Fe + Au, where the angular
distribution of the fragments has been measured \cite{sang}.
The comparison with the QMD calculation is displayed in figure~\ref{gsang}
(in the case of QMD, we have summed over all fragments with $Z\in[5,20]$,
in order to obtain sufficient statistics). 
The angular distribution of the fragment yield turns out to be nicely 
reproduced. The absolute value is also in reasonable agreement. 

Somehow, this overall agreement obtained at the lowest beam energy 
is surprising: In this  regime, at least some of the 
approximations made to derive the QMD equations are not valid anymore. 
These approximations include the quasi particle approximation, 
the neglect of interference effects between subsequent collisions 
(the particles behave like classical billiard balls) 
and the assumption that the Pauli blocking of nucleon scattering can be 
properly 
modeled\footnote{Whereas at high beam energies the few
artificial collisions due to the imperfect Pauli blocking do not
play a significant role, they become increasingly important at low energies
where the number of true collisions decreases rapidly.},...
Nevertheless, what we will retain from these 
comparisons is the increasing evidence that for non-peripheral reactions 
ranging from 30A~MeV to 400A~MeV, the QMD model describes the measured 
data satisfactorily.

\section{Survey of the reaction}
\label{survey-section}
We start our theoretical investigations with a survey of some reactions 
we will encounter later on. 
We illustrate three cases where typical reaction scenarios are expected: 
the participant spectator (PS) scenario (Au~+~Au, 600A~MeV, 8~fm),
the multifragmentation (MF) scenario (Au~+~Au, 150A~MeV, 3~fm) and the 
incomplete fusion (IF) scenario (Xe~+~Sn, 50A~MeV, 3~fm).  
Later, we will discuss in more details to which extent those scenario 
indeed match our numerical simulations. 

\begin{figure}[t]
$$
\epsfbox{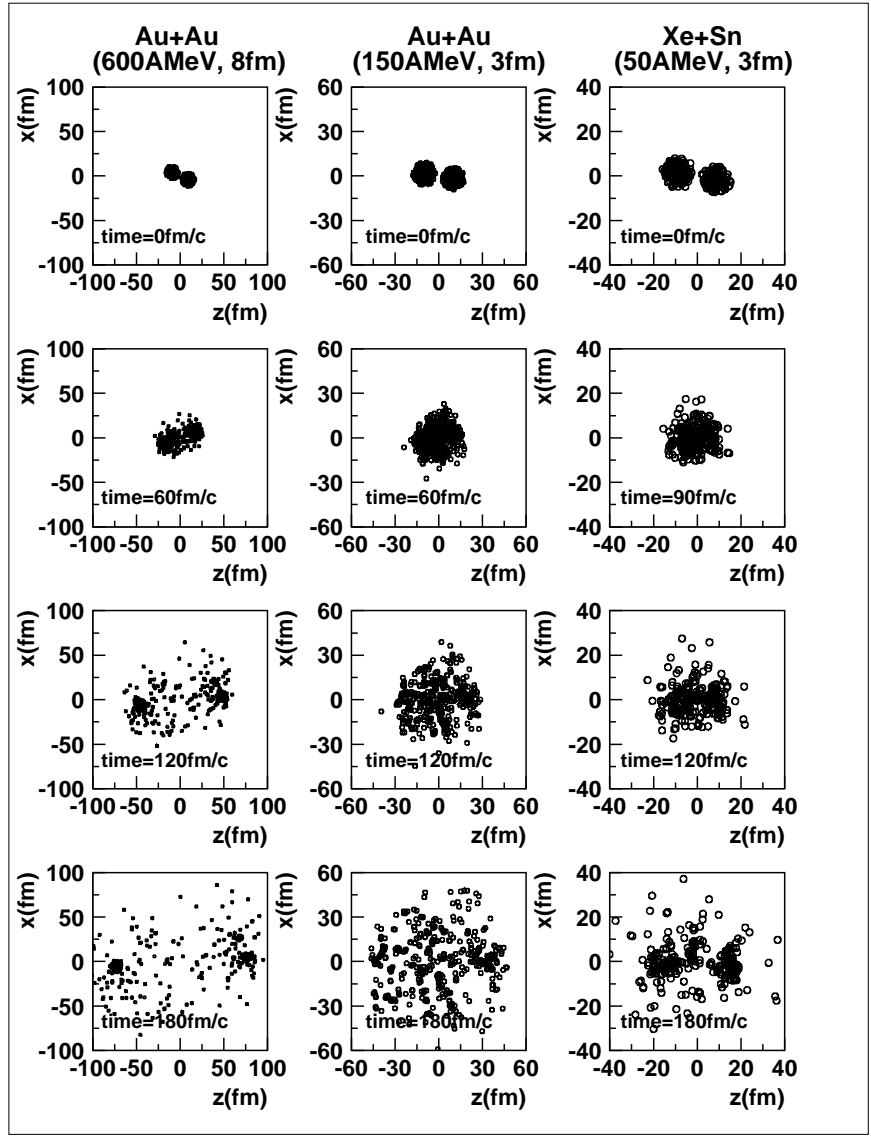}
$$
\caption{Time evolution of nucleon densities (projected on the $xz$ plane) 
for three typical reactions: Au~+~Au, 600A~MeV, 8 fm (left), 
Au~+~Au, 150A~MeV, 3~fm (center) and Xe~+~Sn, 50A~MeV, 3~fm (right).}
\label{nucleons-proj-xz.ps}
\end{figure}

For those three reactions, we select one single event and plot, 
in figure~\ref{nucleons-proj-xz.ps}, the time sequence of the density 
profiles projected onto the $xz$ plane. Each nucleon is marked by a circle 
of 1.5~fm radius. Hence circles which are overlapping belong to the same 
cluster (although the projection may fool a little bit). 
We see three quite different exit channels. The 600A~MeV collision shows 
fragments only around the projectile and target rapidities, as if those 
fragments emerge from the spectator matter. 
At 150A~MeV, we already find fragments at mid-rapidity. The pattern
of particle emission seems isotropic (but future analysis will reveal it 
is not quite so). At 50A~MeV, we observe 
distinct projectile and target-like-fragments as well as mid-rapidity 
fragments which are clearly separated in momentum and coordinate space. 

To conceive plausible scenarios of the reactions, one has to
understand more precisely how the fragments emerge.
In fact, one can already get valuable information by studying 
the evolution of the mean center of mass of all the nucleons 
emerging asymptotically in a given fragment. Let us stress that at 
intermediate times, these nucleons can be very 
dispersed in phase-space. To avoid confusing the reader, we will use the 
word ``prefragment''\footnote{At final time, each prefragment 
indeed becomes a fragment, in the usual acceptation of the word.} 
to denote this ensemble. For each prefragment, let us label 
$A_p$ (resp. $A_t$) the number of nucleons initially contained in 
the projectile (resp. target) and emitted in the asymptotic fragment
\footnote{in the sense of point 2., section~\ref{qmd-model-mult};
sometimes these nucleons will be referred as the prefragment's 
projectile (and target) nucleons}. One has of course $A_{p}+A_{t}=A$, 
the mass of the fragment. Next, we define 
\begin{itemize}
\item 
The center of mass and mean momentum per nucleon of the prefragment as
\begin{equation}
\vec{R}(t)=\frac{\sum_{i=1}^{A} \vec{r_{i}}(t)}{A}\mbox{ and }
\vec{P}(t)=\frac{\sum_{i=1}^{A} \vec{p_{i}}(t)}{A}\;,
\label{equation-cm}
\end{equation}
respectively. In definition~\ref{equation-cm}, $\vec{r}_{i}(t)$ is the 
position of the $i^{th}$ prefragment's nucleon at time t and 
$\vec{p}_{i}(t)$ its momentum.
\item
The center of mass of projectile/target nucleons and
their mean momentum per nucleon:
\begin{equation}
\vec{R}_{p/t}(t)=\frac{\sum_{i_{p/t}=1}^{A_{p/t}} \vec{r}_{i_{p/t}}(t)}
{A_{p/t}}\mbox{ and }
\vec{P}_{p/t}(t)=\frac{\sum_{i_{p/t}=1}^{A_{p/t}} \vec{p}_{i_{p/t}}(t)}
{A_{p/t}}\;, 
\end{equation}
where $\vec{r}_{i_{p/t}}(t)$/$\vec{p}_{i_{p/t}}(t)$ is the 
position/momentum of the $i_{p/t}^{th}$ projectile/target nucleon at time t.
\end{itemize}

\begin{figure}[t]
$$
\epsfbox{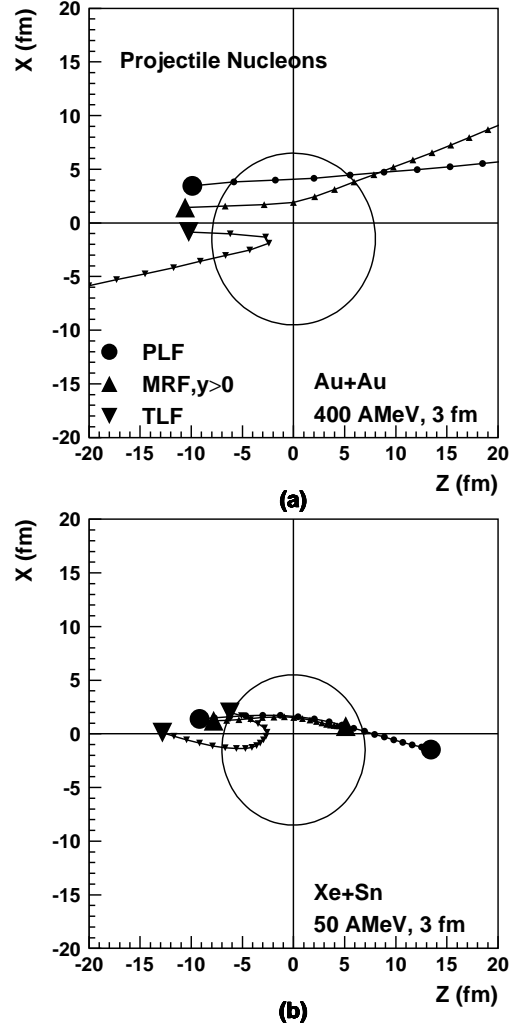}
$$
\caption{Mean trajectories of projectile nucleons ($\overline{X}_{p}(t)$
vs. $\overline{Z}_{p}(t)$) emerging asymptotically in 
projectile-like fragments (PLFs), target-like fragments (TLFs) and
mid-and-positive-rapidity fragments (MRFs, $y>0$) in
Au~+~Au, 400A~MeV, 3~fm (top) and Xe~+~Sn, 50A~MeV, 3~fm (bottom) 
reactions. The initial and final positions are marked by enlarged symbols 
and consecutive time steps (10 fm/c) are connected by a line. The large open 
circles represent the collision partner (target) at the ideal time of 
maximal overlap.}   
\label{rtrajectoriesbis.eps}
\end{figure}

Then, we perform the average (denoted by $\overline{\vphantom{X}...}$)
over all fragments, possibly selecting a given class, and over a 
large ensemble of QMD events.  
From figure~\ref{nucleons-proj-xz.ps}, it is obvious that 3 subclasses of 
heavy fragments have to be privileged:
\begin{itemize}
\item
The heavy fragments emerging around the projectile rapidity, referred 
hereafter as {\em projectile-like-fragments} (PLF).
\item
The heavy fragments emerging around the target 
rapidity, referred hereafter as 
{\em target-like-fragments} (TLF). 
\item
The heavy fragments emerging at mid-rapidity, referred hereafter as 
{\em mid-rapidity fragments} (MRF).
\end{itemize}
In figure~\ref{rtrajectoriesbis.eps}, we present $\overline{X}_{p}(t)$
as a function of $\overline{Z}_{p}(t)$ in Au~+~Au (400A~MeV, 3 fm) 
and Xe~+~Sn (50A~MeV, 3~fm) reactions, for those three subclasses.
In the case of MRFs, we have limited ourselves to those fragments 
emerging with $P_{z}\left(t_{fin}\right)>0$, to avoid trivial cancellations.

At 400A~MeV, we observe strong transverse correlations, in 
agreement with a (refined) PS model: basically,
the projectile nucleons which emerge in PLFs are initially lying 
in the projectile spectator-zone, so avoiding 
the target. Looking more carefully at the data, one notices that these
nucleons have, at initial time, a finite mean momentum in $x$ direction.
We have evaluated $\overline{P_x}^{PLF}(t=0)\approx\mbox{20 MeV/c}$,
which is already an appreciable fraction of the directed flow observed 
experimentally. 

The projectile nucleons emerging in TLFs\footnote{By symmetry, the
same is true for target nucleons emerging in PLFs.} are located on the 
other face of the projectile nucleus, where they will 
be ripped off more easily by the target spectators. They also possess a 
finite transverse momentum, in the $-x$ direction, which helps them 
in that respect.

As expected, MRFs' nucleons are lying in the participant region.   
In the course of the reaction, they are stopped and gain transverse 
momentum by interacting with surrounding nucleons; one may conjecture 
that this is the onset of the blast wave observed recently in very 
central collisions \cite{je}. 

At 50A~MeV, the situation is quite different. Firstly, 
no more spatial transverse correlation is observed. As a matter of
fact, the clear-cut PS model supporting them naturally looses its validity 
at low energy. 

Secondly, the blast has completely disappeared because the dynamics is 
now dominated by the attractive part of the mean-field potential. 
Even if nucleons emerging in PLFs and MRFs are still
momentum-preselected and tend to point out of the overlapping region,
they are bent in the course of the reaction by their strong 
attractive interaction with the nucleons of the other classes; therefore,
we observe ultimately an even larger flow in the $-x$ direction
\footnote{At 150A~MeV, attraction and repulsion balance each other.}.
Hence, the projectile and target have a tendency to rotate
around a common axis; this seems to be the onset of the deep inelastic 
collision (DIC) regime, encountered at higher impact parameters. 
Let us remind that, according to the DIC scenario, MRFs' nucleons 
equilibrate and reside almost exactly at the center of mass, while PLFs' 
nucleons rotate at the periphery and are only emitted once this rotation has 
taken place. 

Nevertheless, figure~\ref{rtrajectoriesbis.eps} clearly proves
that for a beam energy of 50A~MeV and an impact parameter of 3~fm, 
PLFs' projectile nucleons {\em traversed} the whole target nucleus 
more than rotated around it (and vice versa). In fact, they followed the
same trajectory as MRFs' nucleons but went on farther for a reason 
still unclear at this stage---if any reason at all---other than a statistical 
fluctuation. This ``transparency'' must be strongly contrasted with the 
high energy case, where PLFs' nucleons come (almost) exclusively from the 
spectator parts. We will discuss this phenomenon in more details later.

\section{Does the whole system equilibrate?}
\label{whole-therm-sect}
Ideally, one should test the phase space occupation in 
the exit channels to answer this question. However, their number makes 
this ideal test impossible in practice. Therefore, we limit ourselves 
to check whether some necessary criteria are satisfied:

\subsection{Equilibration according to $E_{rat}$}
\label{equil-erat-section}
In a thermalized system, the momenta distribution must be isotropic. 
For a nucleus-nucleus collision, the original longitudinal anisotropy 
has first to be destroyed by nucleon-nucleon interactions. Hence,
an isotropic distribution of emitted matter is not trivially achieved.
For instance, figure~\ref{nucleons-proj-xz.ps} already shows qualitatively 
that the stopping may be only partial.  

\begin{figure}[t]
$$
\epsfbox{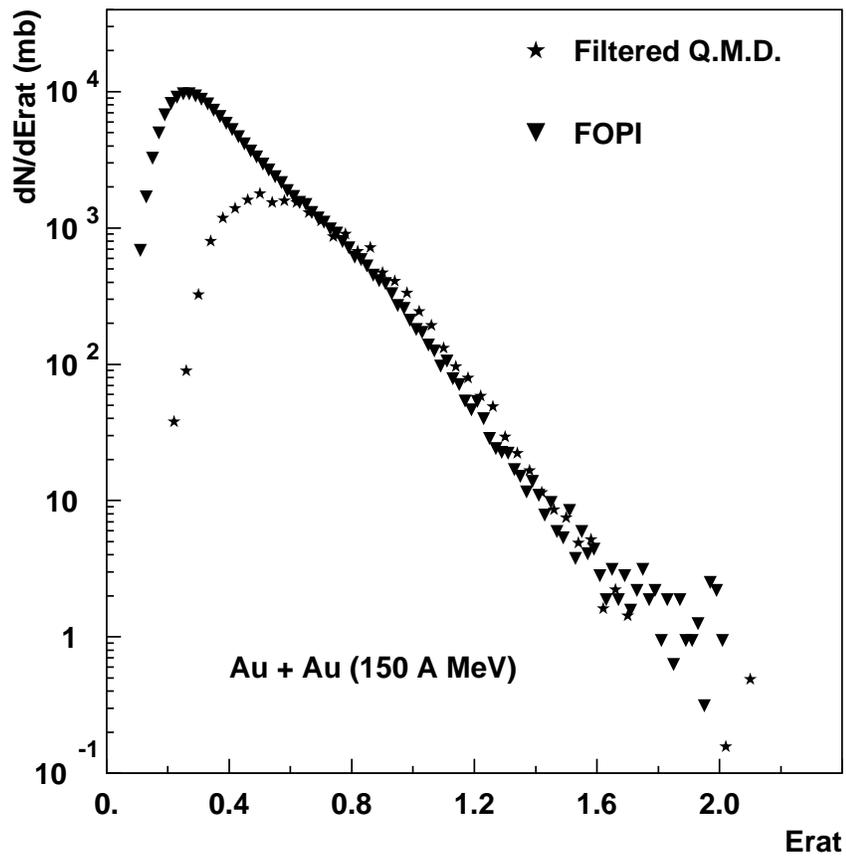}
$$
\caption{$\mbox{d}N/\mbox{d}E_{rat}$ of filtered QMD-simulations 
as compared to the FOPI experiment.}
\label{erats}
\end{figure}

Possible anisotropy can be quantified by the differential cross section 
$d\sigma/dE_{rat}$, where Erat is defined as
\begin{equation}
E_{rat} = {\sum p_{\perp}^2/2m_{N} \over \sum p_{z}^{2}/2m_{N}}.
\end{equation}
Equilibration corresponds to $E_{rat}=2$. This quantity has been 
measured at 150A~MeV by the FOPI collaboration and compared with QMD 
(cf. figure~\ref{erats}). The agreement demonstrates that for central 
collisions (the calculation was stopped at b~=~7~fm)
the excitation energy of the system is properly described
in the QMD model\footnote{The stopping is
a complicated interplay between mean field, collisions, Pauli
blocking...and thus pretty intricate to reproduce.}. 
The acceptance cuts of the detector lower the $E_{rat}$ values
as compared to the QMD events in which all particles are counted. 
However, independently of the acceptance, one can formulate the general 
statement that $E_{rat} = 2$ and global equilibration are achieved only in 
rare events at that energy\footnote{For further comparisons performed by 
the FOPI collaboration, we refer to a recent publication \cite{je}.}
We have evaluated $d\sigma/dE_{rat}$ at lower energies 
(Xe~+~Sn at 32 and 50A~MeV), where data have been taken but not analyzed yet.
At large values of $E_{rat}$, we observe about the same slope 
as for the 150A~MeV reaction. Consequently, $E_{rat}$ values larger than 
1.5 are almost never obtained. 

Of course, one can argue that experimentally, all impact parameters
are mixed, whereas the physics can differ sensibly from small to
large impact parameter. For instance, in the PS model taken at finite 
impact parameter, one may conjecture that the participants equilibrate, 
but the spectators 
certainly do not. A global thermalization is therefore unrealistic. 
On the other hand, for very small $b$, no more nucleon can be considered as a 
spectator and global equilibrium might be achieved. This motivates 
us to investigate the 
dependence of $d\sigma/dE_{rat}$ on $E$ {\em and} $b$. For each fixed 
value of $b$, $d\sigma/dE_{rat}$ turns out to be relatively peaked around 
its mean value ($\overline{E}_{rat}$), displayed in table~\ref{table-erat}, 
for Au~+~Au and Xe~+~Sn reactions at 50, 150, 400A~MeV and 50A~MeV
respectively.

For nearly all energies and impact parameters, $\overline{E}_{rat}\ll 2$.
This discards the global equilibrium hypothesis. Only for E larger than 
$\approx 400$ A~MeV, one might come close to global equilibration in very 
central collisions (which however do not contribute significantly to the 
total cross section). Moreover, if one remembers the dependence of the IMF 
multiplicity on the beam energy (in figure~\ref{imf_nc_msu_comp_part2}),
one can infer that global equilibrium is seldom found 
simultaneously with multifragmentation. 

For all energies, a higher impact parameter results 
in a higher anisotropy of the momentum distribution.
But this is particularly true at high energies, where
the initial transverse position of a nucleon is a strong 
prerequisite to its destiny. At lower 
energies, such a geometrical criteria is not expected,
and the dependence of $\overline{E}_{rat}$ on $b$ is
indeed weaker.

We have just observed and understood that a global 
equilibrium cannot be achieved at large E and large b.
One can wonder why such a conclusion holds at small E
($\approx 50$ A~MeV) at all. This is precisely one of the
concerns of this paper. However, without going much further,
one realizes that at the same energy and same $\hat{b}=b/R$, 
$\overline{E}_{rat}$ is smaller for the Xe~+~Sn system than 
for the Au~+~Au one,... a natural feature in the interaction of 
non-opaque objects.

\subsection{Equilibration according to the momentum distribution 
of specific fragments}
\label{section-dist-final}
In the previous section, it was shown that the momentum 
distribution of emitted particles was globally anisotropic. 
But is it true for each particle species?
For instance, the thermal form of the {\em single particle}
spectra gave rise to speculations that the system may have  
thermalized, in apparent contradiction with what we 
have just deduced.

\begin{figure}[t]
$$
\epsfbox{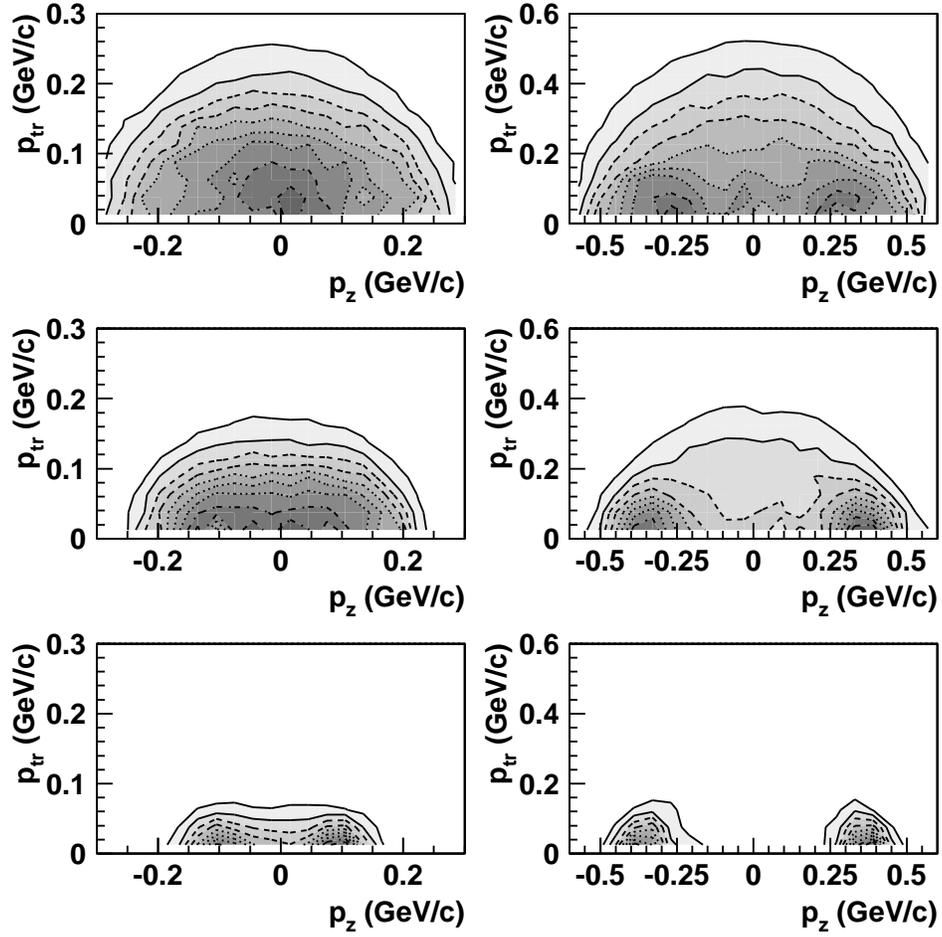}
$$
\caption{$\mbox{d}^{2}\sigma/p_t\mbox{d}p_t\mbox{d}p_z$ for A~=~1 (top), 
$2\le A\le 4$ (middle) and $A\ge 5$ (bottom) produced in Xe~+~Sn, 50A~MeV, 
3~fm (left) and Au~+~Au, 400A~MeV, 3~fm (right) reactions.}
\label{trajectories2.ps}
\end{figure}

To get a better insight, we present (figure~\ref{trajectories2.ps})
the differential production spectra 
($\mbox{d}^2\sigma/p_t \mbox{d}p_t \mbox{d}p_z$) of singles, light 
($2\le A \le4$), and heavy ($5 \le A$) fragments for 
Xe~+~Sn (50A~MeV) and Au~+~Au (400A~MeV) reactions at 
b~=~3~fm \footnote{Even if not displayed, let us mention that 
the Au~+~Au (150A~MeV) reaction nicely interpolates the 50A~MeV and 
400A~MeV ones}. In this presentation, emission from a thermalized 
source produces circles of constant cross section around the source velocity. 

At 50A~MeV, the proton distribution is indeed nearly isotropic
and could be attributed to a thermal source located at mid-rapidity. 
However, the light fragment distribution already contains a 
forward-backward enhancement which definitively evolves to a 2 
sources emission-pattern for the heavy fragment production.
  
The situation at 400A~MeV is somehow different: We observe 3 sources of 
protons that we naturally attribute to the mid-rapidity
fireball, the projectile and the target remnants. 
For heavier fragments, the mid-rapidity source is weaker
and the spectra more and more dominated by fragments 
emitted from projectile and target remnants. In other
words, the importance of the mid-rapidity source decreases with 
the size of the fragment under consideration. 

From figure~\ref{trajectories2.ps}, we can conjecture that 
for a given beam energy, the fragment distribution is {\em always} more 
anisotropic than the proton one. This clarifies the apparent paradox 
presented at the beginning of this section. This also implies that 
fragment spectra should be used preferably to proton ones for 
testing the degree of thermalization of any system.

In the framework of our model, b~=~3~fm collisions never generate globally 
equilibrated nuclear matter. At high energy, it is a more or less 
a trivial consequence of the geometrical cuts encountered
in the PS scenario, as mentioned previously. At low energy, another 
explanation has to be found. To disentangle the role of such cuts from other 
possible causes, we have analyzed the same reactions at b~=~0~fm.
At 400A~MeV, distributions are close to isotropic
whatever\footnote{This is the new insight of as compared to the value 
of 1.67 found in table~\ref{table-erat}} the class considered.
This confirms that most of the anisotropy found at finite $b$ can be 
interpreted inside the PS scenario. 

The situation is quite the opposite at 50A~MeV. There, the distribution 
patterns found at b~=~3~fm and b~=~0~fm are basically identical.
This observation definitively discards explanations exclusively based on 
some relative rotation (which would inhibit a complete fusion
between the two partners, as it is invoked in the case of DIC
scenario).

By now, we have accumulated enough insights to understand
that at 50A~MeV and small impact parameters, the majority
of the projectile nucleons which are finally entrained in fragments 
have {\em traversed} the target (and vice versa). This is only possible
because the Pauli principle blocks almost all collisions at this energy. 
Accordingly, the transparency anticipated at the end of 
section~\ref{survey-section} indeed represents the mean behavior, and not 
a fluctuation on top of it.

Nevertheless, we must still reconcile this transparency hypothesis 
with the isotropy of the proton momentum-distribution.
For this purpose, we use the following feed back argument: 
Initially, the system is both organized 
(nucleon-nucleon correlations) and anisotropic.
We suggest that the nucleons finally emitted as singles are 
precisely the ones which have encountered the most 
violent/numerous collisions, necessary to destroy those 
correlations. 
Accordingly, their momentum distribution
reaches a form compatible with a thermal spectra
pretty fast. On the contrary, final fragments
contain nucleons relatively unaffected by those collisions 
and of course maintain a memory of their initial momenta 
much longer.

\begin{figure}[t]
$$
\epsfbox{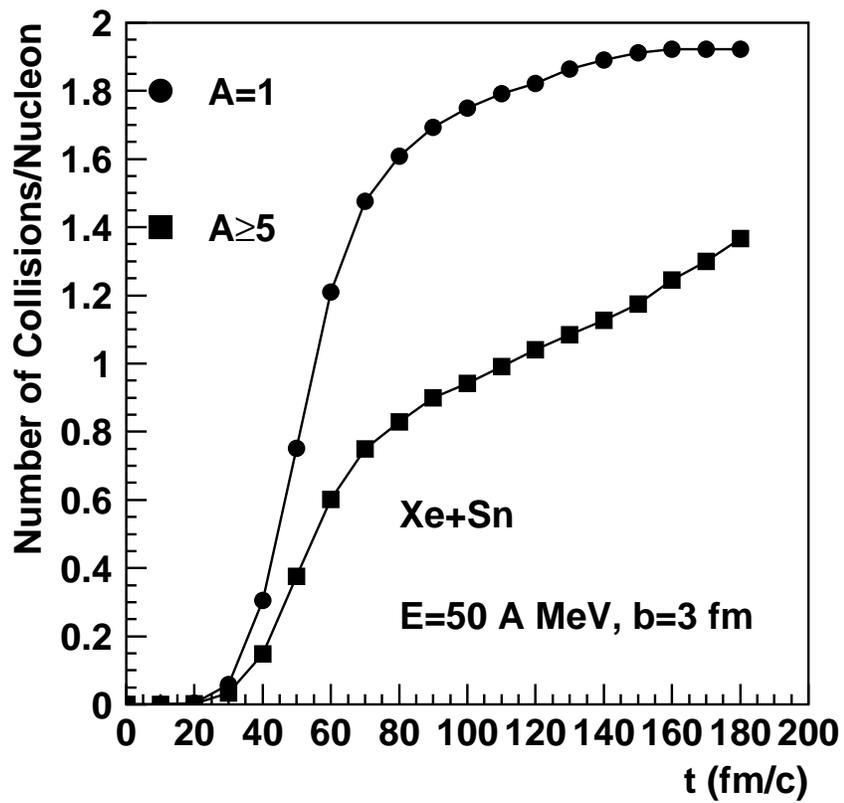}
$$
\caption{mean number of collisions per nucleon for nucleons emerging
asymptotically (a) as singles (b) in $A\ge 5$ fragments. Xe~+~Sn collisions 
at 50A~MeV beam energy and 3~fm impact parameter were considered.}
\label{coll.eps}
\end{figure}

The plausibility of this conjecture is checked in figure~\ref{coll.eps}
where we have displayed, for the
Xe~+~Sn (50A~MeV, 3~fm) reactions, the mean number of collisions 
per nucleon as a function of time. We focus on
those nucleons finally (a) emitted as singles, (b) members of $A\ge 5$ 
fragments. From figure~\ref{coll.eps}, it appears that nucleons which 
turn out to be singles in the final stage have indeed undergone a higher 
number of collisions than those contained in fragments. For large times, the 
hadronic gas becomes more and more dilute, the single nucleons tend to 
evolve freely and the collision rate vanishes. On the other side, however 
isolated a fragment may be, its nucleons still collide with one another:  
This explains the constant collision rate ($\approx 1$ collision 
every 200~fm/c) observed asymptotically. These internal collisions do not 
modify the chemical composition in large respects. Those happening at early 
times are much more important because they determine the later composition. 
Then, to go beyond the qualitative picture, only the collisions happening 
during an effective reaction time of $\approx$~80~fm/c should be taken into 
account. During this effective time, nucleons finally emitted as
singles have collided roughly twice as much as nucleons emitted in 
heavy fragments... a spectacular difference! For the time, we do not pursue 
this analysis any further; we just retain the validity of our feed back 
argument and the incompatibility of results found in figure~\ref{coll.eps} 
with a global-equilibrium hypothesis.

In fact, the previous discussion is a perfect non-trivial example of
time-correlation. It illustrates the power of these time-correlations 
in investigating the dynamics of nucleus-nucleus collisions.
Later, in sections~\ref{init-fin-corr-section} and
\ref{time-evol-of-reac}, we will study some of them more systematically.
But first, we would like to conclude the topic of global equilibration 
by addressing it from another viewpoint: the mixing of projectile and target 
nucleons or, in other words, the ``chemical'' equilibration.

\subsection{Equilibration according to the mixing of projectile and target 
nucleons}
\label{equil-projec-target-section}

\subsubsection{Motivation and redefinition of projectile/target-like
and mid-rapidity fragments}
\label{mrf-plf-def}
We have already mentioned that PLFs, TLFs and MRFs may have a quite 
different origin. Accordingly, it is usually worthwhile to study
these types of fragments separately, as we have already done in
figure~\ref{rtrajectoriesbis.eps}. For this purpose, it is
very convenient to deal with a more practical (and maybe more general)
criteria defining those classes. Such a criteria emerges naturally once
we have realized that at high energy, PLFs (TLFs) are mainly composed of 
projectile (target) nucleons, while MRFs contain projectile and target 
nucleons with identical probability (for symmetric reactions). 

Accordingly, from a theoretical point of view, we obtain more operational 
definitions of PLFs, TLFs and MRFs as following: 
we first remind the reader that $A_{p\,i}$ (resp. $A_{t\,i}$) 
has been defined as the number of nucleons initially contained in the 
projectile (resp. target) and finally emitted in the $i^{\mbox{th}}$
fragment. Next, we define 
\begin{equation} 
a_i = \frac{A_{p\,i}}{A_{p\,i}+A_{t\,i}}=\frac{A_{p\,i}}{A_{i}}
\end{equation}
as the proportion of projectile nucleons in the $i^{\mbox{th}}$
fragment and classify the fragments according to their $a_i$:
\begin{eqnarray}
\mbox{HF with}\hphantom{X}
0.00\le a_i < 0.25 &:& \mbox{target-like-fragment (TLF)}\nonumber\\ 
0.25\le a_i < 0.75 &:& \mbox{mid-rapidity fragment (MRF)}\nonumber\\
0.75\le a_i \le 1.00 &:& \mbox{projectile-like-fragment (PLF)} 
\label{def-PLF-TLF}
\end{eqnarray}

At high energy and finite impact parameter, both definitions 
coincide. Nevertheless, the definition given in formula~\ref{def-PLF-TLF}
is more general (and will be adopted from now on), as it can also 
be used (see \S~below) to investigate chemical equilibration 
even in the absence of three sources clearly separated in rapidity,
for instance at vanishing impact parameter or low energy.
In fact, a detailed study of the rapidity distribution of TLFs, MRFs and 
PLFs has revealed that 
\begin{equation}
y_{\mbox{targ}}\approx
\overline{y_{TLF}}<\overline{y_{MRF}}<\overline{y_{PLF}}
\approx y_{\mbox{proj}}\;,
\end{equation}
whatever the energy and the impact parameter
and however small $y_{\mbox{proj}}-y_{\mbox{targ}}$ may be. 
This explains why we stick to ``mid-rapidity fragment'' as a faithful 
label of the $0.25\le a_i < 0.75$ class, even if the new 
classification~\ref{def-PLF-TLF} does a priori not rely on any criteria
involving the rapidity itself.

If the fragments are formed after the system has passed through
a fully equilibrated phase, as it is expected in the compound nucleus 
scenario, we expect a distribution of $a_i$
according to a binomial law with a mean value of 0.5 
(for symmetric reactions) and a variance inversely proportional to 
the square root of the fragment mass. Thus $a_i$ distribution should be 
sharply peaked around 0.5 for heavy fragments. 
On the other hand, in the simplest version of the PS model,
we expect $a_i$ to be either 0 (TLF) or 1 (PLF). If 
participant and spectator matter interact weakly, we expect two 
narrow peaks around these values. What about the QMD results?

\subsubsection{Distribution of projectile/target-like and mid-rapidity 
fragments}
\begin{figure}[t]
$$
\epsfbox{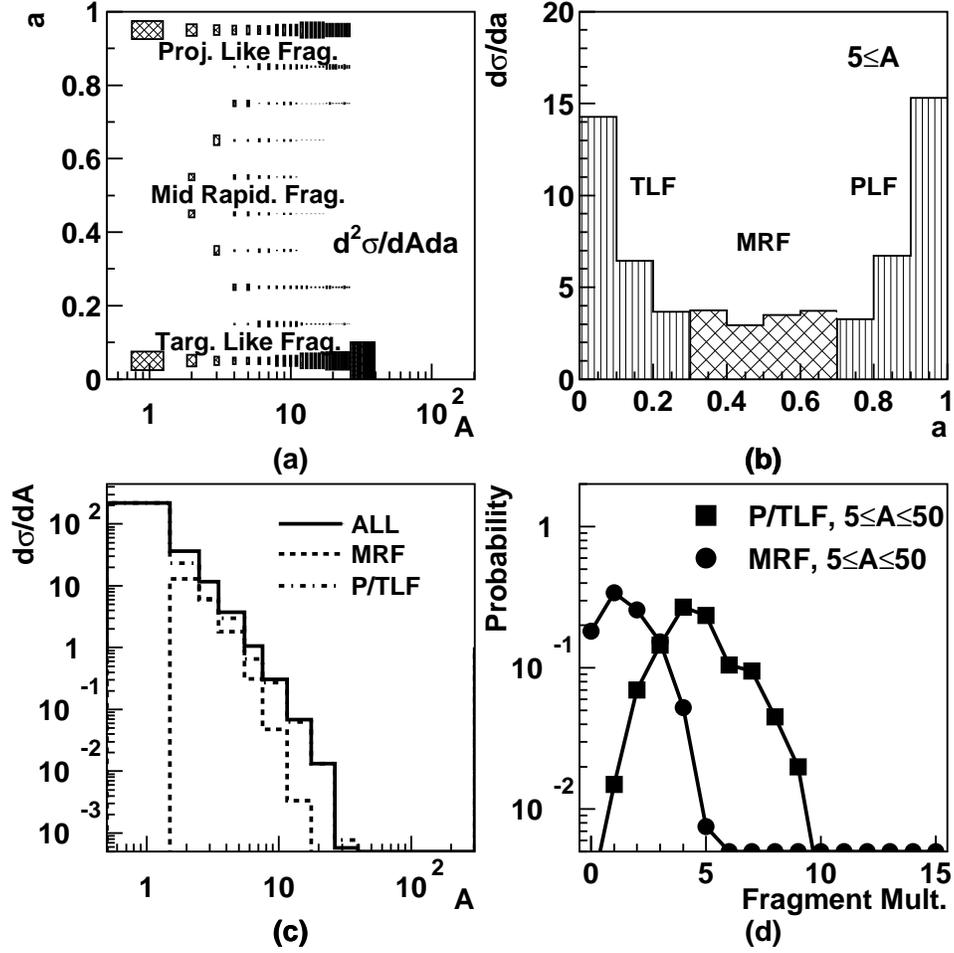}
$$
\caption{P/TLF and MRF distributions for Au~+~Au, 400A~MeV, 3~fm reactions
evaluated in the model (A). In panel (a), we illustrate the (normalized) 
double differential production cross section: $\left(\mbox{d}^2\sigma/
\mbox{d}a\mbox{d}A\right)/\left(\mbox{d}\sigma/\mbox{d}A\right)$, 
where $a$ is the proportion of projectile nucleons in the fragment.
Panel (b) represents $\left.\mbox{d}\sigma/\mbox{d}a\right|_{A\ge 5}$,
panel (c), the mass spectra of PLF ($0.75<a\le 1.$) $\cup$ 
TLF ($0.\le a<0.25$), MRF ($0.25\le a\le 0.75$) and all fragments
taken together. The multiplicity distributions are displayed in panel (d).}
\label{auaue400_b3_diffaprop.ps}
\end{figure}

\begin{figure}[t]
$$
\epsfbox{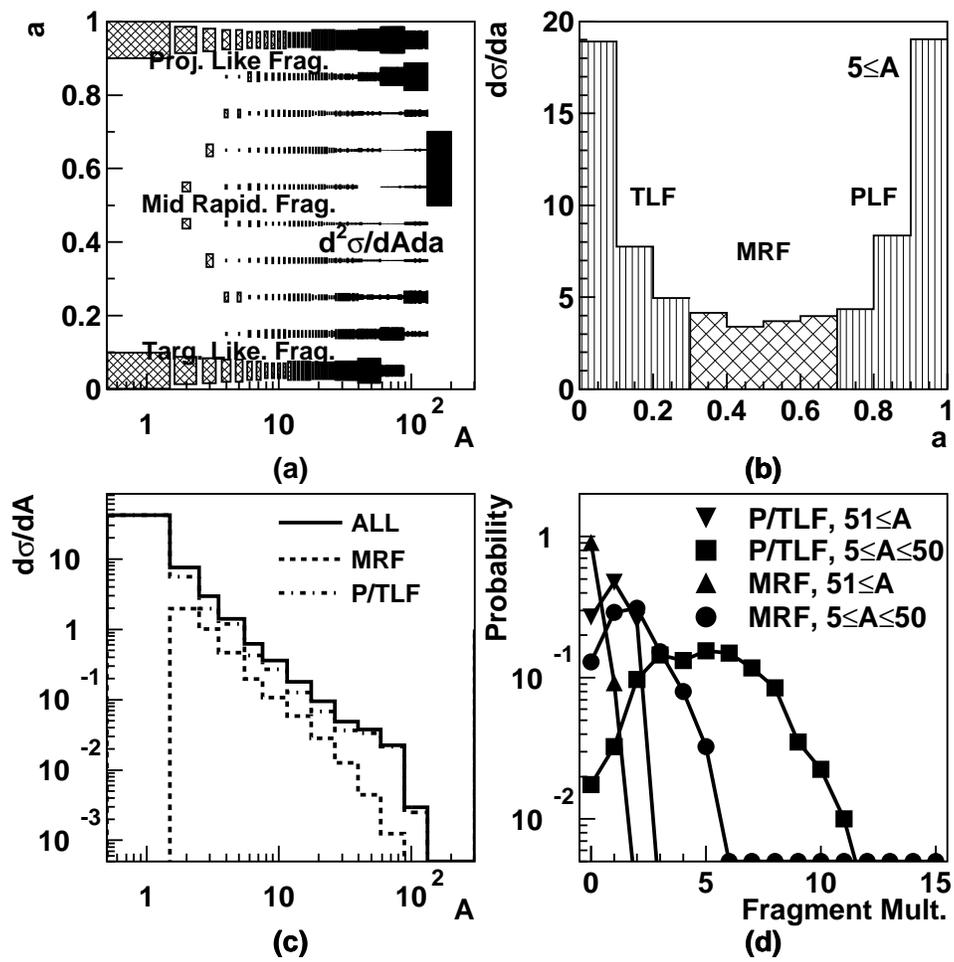}
$$
\caption{same as figure~\ref{auaue400_b3_diffaprop.ps} for 
Xe~+~Sn, 50A~MeV, 3~fm reactions.}
\label{xesne50_b3_diffaprop.ps}
\end{figure}

In figures~\ref{auaue400_b3_diffaprop.ps} and \ref{xesne50_b3_diffaprop.ps}, 
we present results of the model (A) for Au~+~Au, 400A~MeV, 3~fm and 
Xe~+~Sn, 50A~MeV, 3~fm reactions.
In panel (a), we generalize the usual notion of mass spectrum 
$\mbox{d}\sigma/\mbox{d}A$: In fact, more detailed 
information is carried by the (normalized) double differential production 
cross section: 
$(\mbox{d}^2\sigma/\mbox{d}a\mbox{d}A)/(\mbox{d}\sigma/\mbox{d}A)$
\footnote{The normalization factor $\mbox{d}\sigma/\mbox{d}A$  
guarantees that boxes are still visible at large A, where the 
mass spectrum decreases steeply. The price one has to pay for this type
of presentation is the loss of any information about the absolute mass yield 
in this plot.}. In panel (b), we present the differential cross-section for 
the production of HFs: $\left.\mbox{d}\sigma/\mbox{d}a\right|_{A\ge 5}$.
This quantity is normalized to the total HF multiplicity. In panel (c), 
we plot the global mass spectrum, as well as the more specific mass spectra 
of P/TLFs and MRFs. Finally, in panel (d), the fragment-multiplicity 
distribution for IMFs ($5\le A\le 50$) and very heavy fragments 
(VHFs, $51\le A$) are presented (see figures for more details). 

We first discuss the high energy reactions. From the two upper panels
of figure~\ref{auaue400_b3_diffaprop.ps}, we conclude that, independently 
of $A$, the overwhelming majority of fragments are either projectile-like or 
target-like: at 600 A~MeV (not displayed), about 15\% of the fragments 
are MRFs in nearly central collisions ($b_<$) and less than 1\% in peripheral 
collisions ($b_>$). At 400A~MeV these proportions increase to about 30\% ($b_<$) and 4\% ($b_>$). 

Focusing on panel (c), we see 
that for high energy central collisions, the mass spectrum of MRFs and PLFs 
are almost identical in form---a power law. This is a remarkable result in 
view of the different production mechanisms.
It confirms the conjecture that the mass spectrum is not very sensitive to
the underlying physical process... a conclusion almost unavoidable
if one remembers that the experimental mass spectrum is a pretty
robust function of the system considered. Let us just mention that in 
high energy {\em peripheral} collisions, the MRF and P/TLF mass spectra 
naturally depart from one another: while no MRF of mass larger than 8 is 
found, the P/TLF spectrum extends up to A~=~150, with a local minimum 
around A~=~30-40.

From panel (d), we see that on the average 6 IMFs are produced 
(5 P/TLFs and 1 MRF), but no VHF. From the whole 
figure~\ref{auaue400_b3_diffaprop.ps}, one concludes that 
even though no heavy remnant is observed at such a low impact 
parameter, the memory of the entrance channel is fairly preserved. 
In the peripheral reaction already mentioned, two heavy remnants 
usually survive, in concordance with a ``$<$'' shape of the double 
differential cross section, but about 2 P/TLFs are produced as well. 

In peripheral collisions, the typical behaviors observed at 50A~MeV are 
pretty similar to those just discussed at 400A~MeV (for $b_>$). Even though 
the total number of MRFs is larger at 50A~MeV, it still does not represent 
more than 20\% of all emitted HFs. As for the average fragment multiplicity, 
2 VHFs survive the reaction and about 5 IMFs are produced (4 P/TLFs and 1 MRF)
at low energy. In fact, the major discrepancy with the high energy case 
occurs in the study of the P/TLF and MRF mass spectra: these are now 
similar, up to A~=~20. Also, P/TLFs of mass $A\approx 30-40$ are produced 
more abundantly and tend to fill the valley encountered at higher energy.

In low-energy central collisions (illustrated in 
figure~\ref{xesne50_b3_diffaprop.ps}), about 25\% of all fragments are 
MRFs, while both mass spectra (MRF and P/TLF) are very close in form for 
almost all values of A. On the average, we do not observe two but only one 
remnant (VHF) which is {\em either} projectile {\em or} target-like, in 
definitive contradiction with a compound nucleus/incomplete fusion scenario. 
The disassembling of the other remnant is the basic cause of the broad 
mass-spectrum and the source of large fluctuations in the IMF yield. 

\begin{figure}[t]
$$
\epsfbox{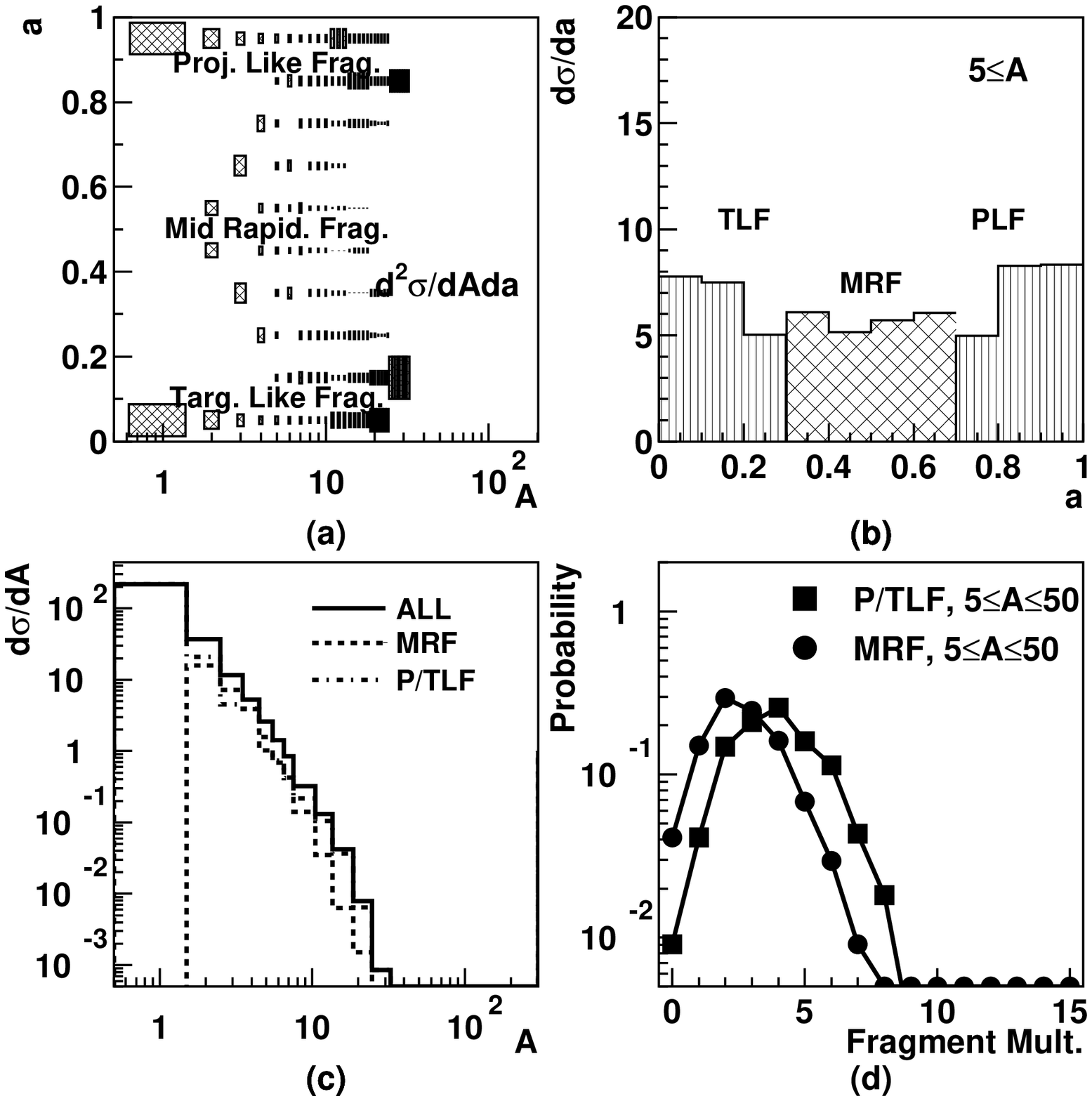}
$$
\caption{same as figure~\ref{auaue400_b3_diffaprop.ps} with model (B).}
\label{auaue400_b3flip_diffaprop.ps}
\end{figure}

\begin{figure}[t]
$$
\epsfbox{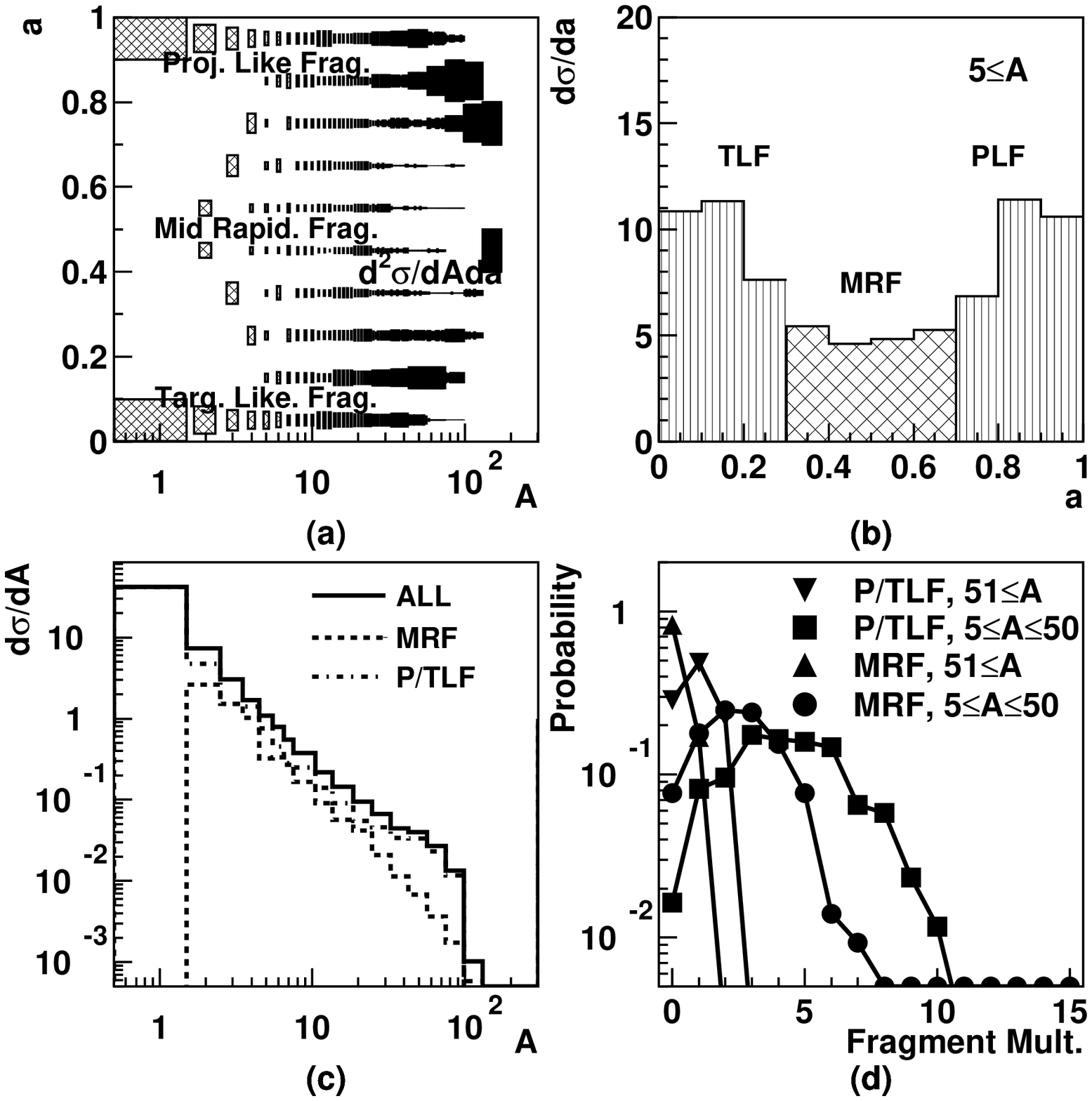}
$$
\caption{same as figure~\ref{xesne50_b3_diffaprop.ps} with model (B).}
\label{xesne50_b3flip_diffaprop.ps}
\end{figure}

By studying other energies between 400 and 50A~MeV we have discovered 
nothing but a quite smooth crossover... and therefore established the 
generic character of figures~\ref{auaue400_b3_diffaprop.ps} and 
\ref{xesne50_b3_diffaprop.ps}, at least within the model (A)! 
By definition, results obtained
in the two previous sections (\ref{equil-erat-section} and
\ref{section-dist-final}) are independent of the symmetrized
character of the NN cross section.  
When we come to the question of chemical equilibration
this is no longer the case. Accordingly, we have re-evaluated 
all quantities involved in figures~\ref{auaue400_b3_diffaprop.ps} and 
\ref{xesne50_b3_diffaprop.ps} inside model (B) and
re-displayed them in figures~\ref{auaue400_b3flip_diffaprop.ps} 
and \ref{xesne50_b3flip_diffaprop.ps} respectively.
 
The most spectacular effect is observed for 
$\left.\mbox{d}\sigma/\mbox{d}a\right|_{A\ge 5}$ in panel (b): 
the peaks formerly observed around $a=0$ and 1 are now eroded, while 
at intermediate $a$, the valley is somewhat filled in...
but we are still far from a dominating peak around $a=0.5$,
as it would appear in case of chemical equilibration.
In concordance with the flattening of
$\left.\mbox{d}\sigma/\mbox{d}a\right|_{A\ge 5}$, one observes, in 
panel (d), an average increase (resp. decrease) of the 
MRF (resp. P/TLF) multiplicity by one unit. However, the VHF 
multiplicities are unaffected, as well as the general structure of the 
double differential cross section and the mass spectra (panels (a) and (c)).

On the qualitative level, we can conclude that chemical 
equilibration is achieved neither in model (A) nor in model (B), 
even if one gets closer with this second version of QMD. 

To understand more quantitatively how the chemical equilibration depends 
on the model chosen and on the physical parameters (beam energy, impact 
parameter,...), we provide in table~\ref{chemical-equil} the fraction of MRFs 
among HFs, for both models and various systems (those already chosen in 
table~\ref{table-erat}).
 
Within a wide range of physical parameters, the system is never close to 
chemical equilibrium; indeed, we observe very strong correlations 
between the origin of the particles (in this case projectile {\em or} target) 
and the type of fragment to which they will ultimately belong. In fact,
the majority of fragments ''remain'' projectile-like or target-like.
Mid-rapidity fragments represent more than 40\% of the yield only on 
a band located at high energy and low impact parameter.

In table~\ref{table-dep-dable}, we provide a non-exhaustive list of
qualitative dependences embedded in table~\ref{chemical-equil}.
Let us comment on the third line and clarify the fifth one.

$\bullet$
Selecting b~=~0~fm, and decreasing the beam energy, 
we expect to enter the realm of the quasi-fusion mode, where the
proportion of MRFs should in principle re-increase. At 50A~MeV, this is not 
yet the case.

$\bullet$
Whatever the system, the energy and the impact parameter, it turns out that
the MRF yield inside model (B) is approximatively $3/2$ of the MRF yield 
inside model (A):
\begin{equation}
\sigma_{prod}(MRF)|_{\theta_{cm}\le \pi}\approx\frac{3}{2}
\sigma_{prod}(MRF)|_{\theta_{cm}\le \frac{\pi}{2}}\;!
\label{propor-equil-2mod}
\end{equation}
Of course, this result cannot be universal: a priori, there must 
exist an energy for which a nucleus-nucleus collision happening at b~=~0~fm 
engenders a highly equilibrated phase, whatever the model chosen. However, 
in the energy range relevant for (multi)fragmentation, this is not the case, 
and relation~\ref{propor-equil-2mod} seems to hold. We propose the following 
interpretation: at the early stages of a reaction or in case of a 
poor equilibration, the production of chemically-equilibrated matter is 
proportional to the rate of projectile nucleons
deflected in the phase space of target nucleons (plus a symmetric 
contribution). This deflection rate itself is proportional to the collision 
rate multiplied by the mean deflecting angle associated with the microscopic 
interaction, defined as
\begin{equation}
\bar{\theta}_{deflect}=\mbox{arccos}\left[\int\cos\theta
\frac{1}{\sigma}\frac{\partial \sigma}{\partial \Omega} 
\mbox{d}\Omega\right]
\end{equation}
Precisely, what distinguishes model (A) from model (B) is 
just that interaction. At low energy, 
\begin{eqnarray}
\frac{\partial \sigma}{\partial \Omega}&=&
\frac{\sigma_{NN}}{2\pi}\hphantom{0}\mbox{for $\theta\in[0,\frac{\pi}{2}]$}
\nonumber\\
&&0\hphantom{\frac{\sigma_{NN}}{2\pi}}\mbox{for $\theta\in]
\frac{\pi}{2},\pi]$}\nonumber
\end{eqnarray}
in model (A) and
\begin{eqnarray}
\frac{\partial \sigma}{\partial \Omega}&=&
\frac{\sigma_{NN}}{4\pi}\hphantom{0}\mbox{for $\theta\in[0,\pi]$}\;.
\end{eqnarray}
in model (B). Accordingly, 
\begin{equation} 
\bar{\theta}_{deflectA}(E\ll)\approx \frac{\pi}{3}\mbox{ and }
\bar{\theta}_{deflectB}(E\ll)\approx \frac{\pi}{2}\;.
\end{equation}
In the linear approximation, this permits to reproduce 
relation~\ref{propor-equil-2mod}. If the linear approximation breaks 
down, the ratio $\theta_{deflectB}/\theta_{deflectA}=\frac{3}{2}$ probably 
acts as a scaling factor. From this analysis, we conjecture that, 
quite generally, using model (B) instead of model (A) reduces the 
amplitude of time-correlations by a mere factor, but does not destroy them 
dramatically, as it could have been feared a priori. In the last analysis, 
this would not be so surprising: The (chemical) relaxation times of model 
(A) and (B) seem to differ by just a factor $\approx 1.5$! If global 
equilibration were achieved in one of them, it would be automatically 
achieved in the other. In other words, to choose one model or the other 
would result in the same {\em qualitative} understanding of the underlying 
physics, which is, after all, the main goal of this work.

We now close this whole section devoted to the question
of global equilibration. In fact, table~\ref{table-erat} and 
table~\ref{chemical-equil} reflect two aspects 
of this problem: the equilibration of the momentum distribution 
and the chemical equilibration... They exhibit the same qualitative
dependences on the physical parameters and lead to the same
conclusion: For the impact parameters and energy range (50-600A~MeV) 
investigated in QMD simulations, the system only approaches 
global equilibrium on a domain located at 
very high beam energy and very small impact parameter... a range of no 
interest for multifragmentation.

\section{Initial-final state correlations}
\label{init-fin-corr-section}
In this section, we search for the possible phase-space correlations 
between the initial and the final state of our simulations. To be more 
specific, we would like to know whether nucleons finally entrained in 
fragments were predominantly located in certain regions of phase-space
at the initial time. We shall proceed in several steps.

After detailing the actual way those time-correlations were evaluated
(section~\ref{detail-time-corr}), we investigate separately the case of 
light fragments or singles (LF) and of heavy fragments (HF) in the 
coordinate space (section~\ref{in-fin-correl-coord-space}).
Later (section \ref{in-fin-correl-coord-space-mrf-plf}), we refine 
the HF analysis by focusing on the mid-rapidity fragments and the 
projectile/target-like fragments. 

In a second step, we perform the equivalent study in the 
momentum space (sections~\ref{in-fin-correl-moment-space} and 
\ref{in-fin-correl-moment-space-mrf-plf}). This will help us to 
understand better the question of {\em longitudinal correlation}, 
left open up to there. In section~\ref{catapult-section}, this
problem will clarified by introducing the concept of
``catapult mechanism''. 

\subsection{Evaluation of the time-correlations}
\label{detail-time-corr}
For instance\footnote{This procedure naturally extends to any other 
type of time-correlation.}, suppose we want to know from which part of the 
initial (one-body) phase-space stem the nucleons found in final fragments. 
In the QMD model, this can be done via the following procedure:
\begin{itemize}
\item
  store the initial positions $r_i(t=0)$
  and momenta $p_i(t=0)$ of all $A_P + A_T$ nucleons; 
\item
  simulate the reaction up to $t_{fin}$ (i.e. 180~fm/c);
\item
  perform clusterization at $t_{fin}$ using a minimum spanning-tree
  algorithm with a given clusterization radius (i.e. 4~fm);
\item 
  define $inclass_i = 1$ if at $t_{fin}$ the nucleon i is part of 
  a fragment belonging to the class to investigate. Otherwise
  $inclass_i = 0$
\item
  project the initial positions $ r_i(t=0)$ on a 2 dimensional grid. 
  Each vector $r_i (t=0)$ now corresponds to a grid cell k. 
\item
  define 2 quantities $\rho_{class,k} = \sum inclass_i$ and 
   $\rho_{tot,k} = \sum 1$ where
  the sum runs over all nucleons whose coordinate vector
  falls into the grid cell k
\item
  perform the mean of $\rho$'s on an ensemble of simulations
\item
  display the quantities $\rho_{class,k}$ and 
    $\hat{\rho}_{class,k}=\rho_{class,k}/\rho_{tot,k}$.
\end{itemize}

In case of global equilibration, correlations between 
initial and final states are absent and nucleons emerge equiprobably 
from the initial phase space. It results that $\hat{\rho}_{class,k}$ 
is constant (i.e. independent of $k$), whatever the class. 
On the other hand, in the PS scenario, $\hat{\rho}_{HF,k}$ is 
close to 1 outside the geometrical overlap and close to 0 otherwise.
Of course, the actual physical processes turn out to be more subtle than 
those two idealized pictures, as we will now discuss.

\subsection{Coordinate-space correlations of light and heavy fragments.}
\label{in-fin-correl-coord-space}
\subsubsection{Energy dependence}
We start the analysis of initial-final state correlations
focusing on the coordinate space. 
The left panels of figure~\ref{xz_np.eps} illustrate these correlations for 
nucleons finally entrained in HFs, for several systems: from top to 
bottom: (Au~+~Au, 600A~MeV, 8~fm), (Au~+~Au, 150A~MeV, 8~fm),
(Au~+~Au, 150A~MeV, 3~fm) and (Xe~+~Sn, 50A~MeV, 3~fm).
The shadings correspond to $\rho_{HF,k}$, whereas the boxes size represents 
$\hat{\rho}_{HF,k}$. Both quantities are 
normalized relatively to the grid cell containing the largest value. 
The right panels are just the counterpart for the LF case.

At high energy, the PS model
should apply. As a matter of fact, in panels (a) and (b), we observe 
that HF nucleons come predominantly from the spectator regions,
while LF ones, from the participant zone\footnote{Although not displayed,
this result holds whatever the impact parameter}. 
The transition from the ``participants'' to ``spectators'' is quite 
clear but not as sharp as in the PS model itself. 
This is not the only difference: A closer look at panels (a) and (b) 
reveals that the transverse extent of the ``participants'' and ``spectators'' 
zones depends on z. This longitudinal dependence is absent of the PS
model and we conjecture that it reflects the expansion of the fireball 
already formed by the participants on the front of the nuclei. Due to 
high temperature, these nucleons disassemble very quickly (before the 
projectile and target back ends arrive at the interaction zone) penetrate
the incoming spectator matter and excite it locally (within a mean free path
distance). As a direct consequence, this excited spectator 
matter will form more LFs and less HFs. 
 
At lower beam energy, we observe a gradual disappearance 
of the correlations predicted by the PS model. 
Already at 400A~MeV (not displayed), the correlations are weakened.
Between 400A~MeV and 50A~MeV, the reaction mechanisms 
change completely. At 150A~MeV, 8~fm, we observe 
some spatial correlation only in the distribution of nucleons
emerging as LF (panel (d)). For an impact parameter of 3~fm,
almost no correlation is seen, as it would happen if the system 
were completely equilibrated. However, we have already seen 
in section~\ref{whole-therm-sect} that this hypothesis is not 
correct. 

For nearly central collisions at 50A~MeV, we observe correlations 
along the beam axis. Nucleons at the back ends of projectile and target have 
a higher probability to form a light fragment or to escape as singles than 
those at the front end. With the data at hand, it is not possible to 
interpret this result unambiguously. One could think that the two nuclei 
rotate around each other. Both front ends would fuse in this 
rotation, while the back ends would be ejected by the centrifugal force...
however, we have already seen (see for instance 
section~\ref{survey-section} and more precisely 
figure~\ref{rtrajectoriesbis.eps}) that the rotational character is
rather small at this impact parameter. In fact, what is really happening 
is totally different: 
In section~\ref{catapult-section}, we will explore other
time-correlations and demonstrate that the dense zone formed in the 
projectile-target overlapping region acts as a catapult for the nucleons 
which are initially at the back ends of projectile and target. These 
nucleons are accelerated mostly expelled as singles.

Summarizing, we see a complete change of the coordinate-space 
correlations pattern if we decrease the energy from 600A~MeV to 50A~MeV.

\subsubsection{Quantification of the correlations}
To facilitate the comparison between different energies, impact 
parameters, etc., it is helpful to define a single numerical quantity 
reflecting the importance of the initial-final state correlations. 
For light fragments, we have found appropriate to define 
the ``coordinate space LF correlation number'' as follows:
\begin{equation}
C^{r}_{LF}=\frac{\sqrt{\overline{\left(
\hat{\rho}_{LF}-\overline{\hat{\rho}_{LF}}\right)^{2}}^
{\vphantom{l}}}}{\overline{\hat{\rho}_{LF}}}\mbox{, with }
\hat{\rho}_{LF,k}=\frac{\rho_{LF,k}}{\rho_{tot,k}}\;,
\end{equation}
where $\rho_{LF}$ is the initial density of nucleons 
emerging asymptotically as light fragments and
$\rho_{tot}$ is the mean density in initial 
nuclei. Here, ``$\overline{\vphantom{x}...}$'' represents the average 
taken on all cells $k$ such that 
$\rho_{tot,k}\geq \mbox{max}(\rho_{tot})/10$. 
This condition is set to discard the unphysical fluctuations in the low
density regions. We define $C^{r}_{HF}$, the ``coordinate space HF 
correlation number'', in a quite similar way. 
As $\hat{\rho}_{LF}\leq 1$, we easily establish 
that
\begin{equation}
C^{r}_{LF}\leq\sqrt{\overline{\hat{\rho}_{LF}}^{~-1}-1},
\label{limit-cor}
\end{equation}
(and similarly for $C^{r}_{HF}$),
indicating that correlations are intrinsically limited 
if a large number of nucleons are requested in a given final channel 
(as a limit, they vanish if all the nucleons are needed). 
This is for instance the case for the production of LF in high-energy
central collisions.
Moreover, as $\hat{\rho}_{LF}+\hat{\rho}_{HF}=1$, one has
\begin{equation}
  \overline{\hat{\rho}_{LF}}*C^{r}_{LF}=
  \overline{\hat{\rho}_{HF}}*C^{r}_{HF}
\end{equation}
and of course
\begin{equation}
  \overline{\hat{\rho}_{LF}}+\overline{\hat{\rho}_{HF}}=1\;.
\end{equation}
In table~\ref{table-meanrho} we summarize the values of 
$\overline{\hat{\rho}_{LF}}$ and $\overline{\hat{\rho}_{HF}}$
for Au~+~Au (600, 400 and 150A~MeV) and Xe~+~Sn (50A~MeV) reactions,
at small ($b_{<}$) and large ($b_{>}$) impact parameter. 
In table~\ref{table-correl-spac},
we summarize the values of $C^{r}_{LF}$ and $C^{r}_{HF}$ for the same 
reactions. In high-energy central collisions, one obtains a large number 
of nucleons going to LFs and, according to relation~\ref{limit-cor}, a low 
value of the correlation number. In all other high energy reactions and
fragment classes, strong correlations are present, in agreement with the 
PS model. 

When the energy decreases, more and more nucleons are released 
in HFs and the associated correlation numbers vanish... once again a 
consequence of inequality~\ref{limit-cor}! In fact, at low energy, 
one intuitively expects a transition into the direction of a compound-nucleus 
reaction. This would 
imply a complete mixing of all nucleons and the disappearance of {\em all} 
initial-final state correlations. However, this is clearly not observed:
At large impact parameter ($b_>$), the correlations in $\rho_{LF}$
survive, while for central collisions ($b_<$), they even increase
\footnote{This is now possible due to the relaxation of 
constrain~\ref{limit-cor}. Indeed, less and less nucleons
emerge in LFs at low energy} as compared to the high energy case. 

From table~\ref{table-correl-spac} it also appears that when the 
energy is reduced, the quantities at hand (here, the correlation number) 
depend on the impact parameter in a much smoother way; a fact already 
observed many times, for instance in table~\ref{table-erat}.

\subsection{Coordinate-space correlations of projectile/target-like and
mid-rapidity fragments}
\label{in-fin-correl-coord-space-mrf-plf}
We start the comparison by extracting, in 
table~\ref{mean-rho-part2}, $\overline{\hat{\rho}_{MRF}}$ and 
$\overline{\hat{\rho}_{P\cup LF}}$ for a few systems and both models 
(but we first concentrate our discussion on model (A)).

Table~\ref{mean-rho-part2} confirms that the
PLFs and TLFs outnumber the MRFs at all energies, but this is 
especially true for high energy and high impact parameter. 
$C^{r}_{MRF}$ and $C^{r}_{P\cup TLF}$ are defined as $C^{r}_{HF}$, 
but with the proper selection on the fragment type. These 
correlations numbers are displayed in table~\ref{space-correl-part2}. 

Due to the dominance of PLF and TLF channels, $C^{r}_{P\cup TLF}$ and 
$C^{r}_{HF}$ are pretty close (cf. table~\ref{table-correl-spac}).
More astonishing are the strong correlations exhibited in the MRF channel. 
These are the strongest correlations observed so far.
In central collisions, they even increase when energy is reduced.

\begin{figure}[t]
$$
\epsfbox{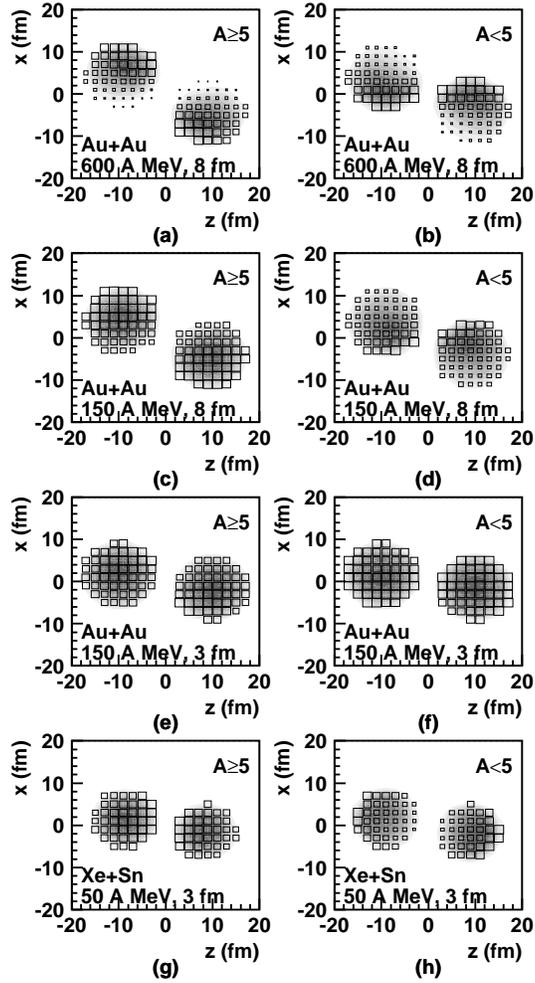}
$$
\caption{Initial-final state correlations in coordinate space:
On the left (resp. right), initial densities of nucleons finally 
entrained in $A\ge 5$ (resp. $A<5$) fragments.
The shading represents those absolute densities , while the area of the 
boxes represents their local ratio with the total density of nucleons.
From top to bottom: (Au~+~Au, 600A~MeV, 8~fm), (Au~+~Au, 150A~MeV, 8~fm), 
(Au~+~Au, 150A~MeV, 3~fm) and (Xe~+~Sn, 50A~MeV, 3~fm).}
\label{xz_np.eps}
\end{figure}

\begin{figure}[t]
$$
\epsfbox{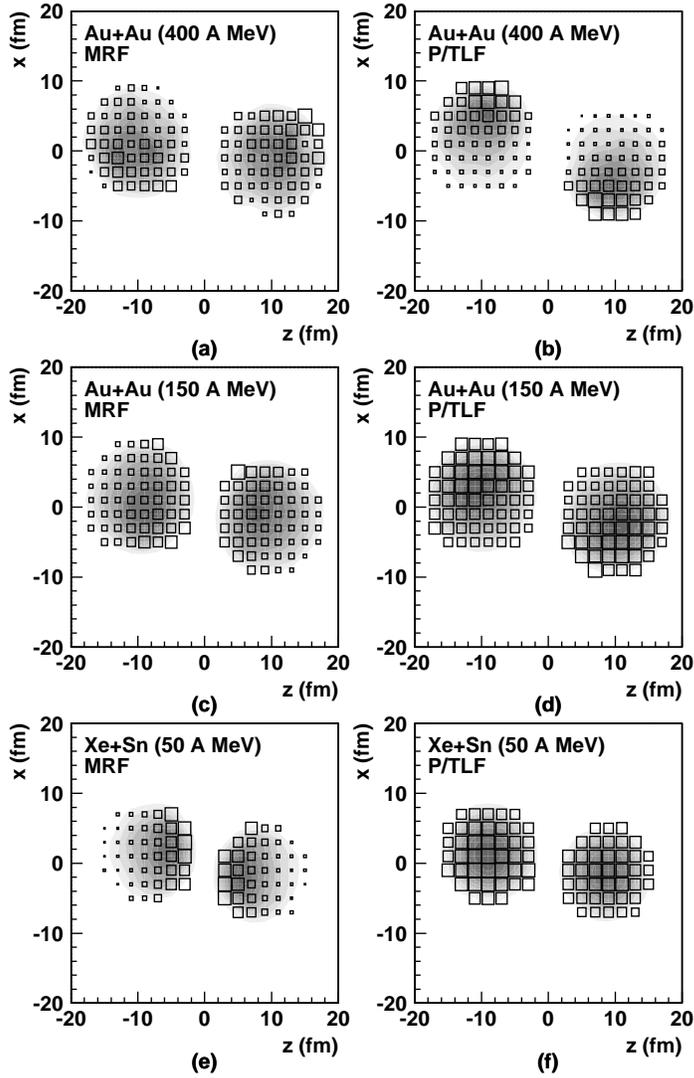}
$$
\caption{Coordinate-space initial-final state correlations:
initial density of nucleons entrained finally in heavy fragments 
$(A\ge 5)$, separately for MRF (left) and P/TLF (right). Three
cases of central (b~=~3~fm) collisions have been considered:
Au~+~Au, 400A~MeV (top), Au~+~Au, 150A~MeV (middle) and
Xe~+~Sn, 50A~MeV (bottom).}
\label{xz_lmf_2p_sb.ps}
\end{figure}

To investigate the nature of these correlations in more details,
we display in figure~\ref{xz_lmf_2p_sb.ps} the coordinate-space 
correlations in the b~=~3~fm reactions\footnote{The peripheral 
reactions qualitatively exhibit the same type of correlations.} computed in 
model (A), following the conventions of figure~\ref{xz_np.eps}. 
Correlations for MRFs and P/TLFs are shown on the left and right side
respectively.

At 400A~MeV, MRFs' nucleons are initially located in the ``participant''
zones, while the PLFs and TLFs' nucleons are found in the ``spectator'' 
zones. These zones are clearly separated: participant nucleons 
penetrate the spectator matter only occasionally, so that its 
composition is basically unchanged. On the other hand, they create and
enter a ``fireball'' which emits a few MRFs in a much more statistical way. 

At 150A~MeV, the PS model does not apply anymore
and the transverse correlations associated with it have indeed completely 
disappeared. This break down (of the transverse time-correlations) was first 
mentioned in section~\ref{in-fin-correl-coord-space} by the inspection of 
figure~\ref{xz_np.eps}. From the panel (e) of this figure, one could 
have been tempted to conclude the absence of (coordinate-space) 
time-correlations in the HF formation process at 150A~MeV. However, the
refined analysis performed in figure~\ref{xz_lmf_2p_sb.ps} reveals that 
the HF fluid is in fact made of three components (MRF, PLF and TLF), 
each of these exhibiting strong time-correlations. This behavior is even 
more pronounced at 50A~MeV, where the nucleons finally entrained in MRFs are 
strongly localized along the longitudinal axis. Why longitudinal? We do not 
have enough information to answer this question right now
\footnote{A possible---but incorrect---scenario 
would be the following: 150A~MeV is not a sufficient energy to create a 
participant fireball over the whole extent of the participant region. 
Rather, a hot spot mixing the nucleons is formed {\em only} around the 
impact point, at $z=0$. For larger values of $|z|$, the temperature 
diminishes, and the nucleons conserve their ``chemical memory''. At later 
stages of the reaction, back ends of both nuclei avoid mixing by rotating 
around the localized hot spot. However, analyzing the temperature profile,
we have found no such a hot spot localized at the contact point, 
but pretty extended isotherms! Moreover, purely rotational 
scenario have already been discarded, for instance by 
figure~\ref{rtrajectoriesbis.eps}}
and leave it open until section~\ref{catapult-section}.

How do the results of model (B) match those of model (A)?
Inspecting table~\ref{space-correl-part2}, MRF correlation numbers
turn out to be reduced by 33\% on the average and by 50\% at the most 
(for 150A~MeV). PLF correlation numbers are basically left untouched, 
whatever the physical parameters. Even if not displayed here for the 
purpose of concision, let us stress that when correlations illustrated 
in figure~\ref{xz_lmf_2p_sb.ps} are re-evaluated inside model (B), they 
present exactly the same characteristic pattern, slightly attenuated.

\subsection{Momentum-space correlations of light and heavy fragments}
\label{in-fin-correl-moment-space}
\begin{figure}[t]
$$
\epsfbox{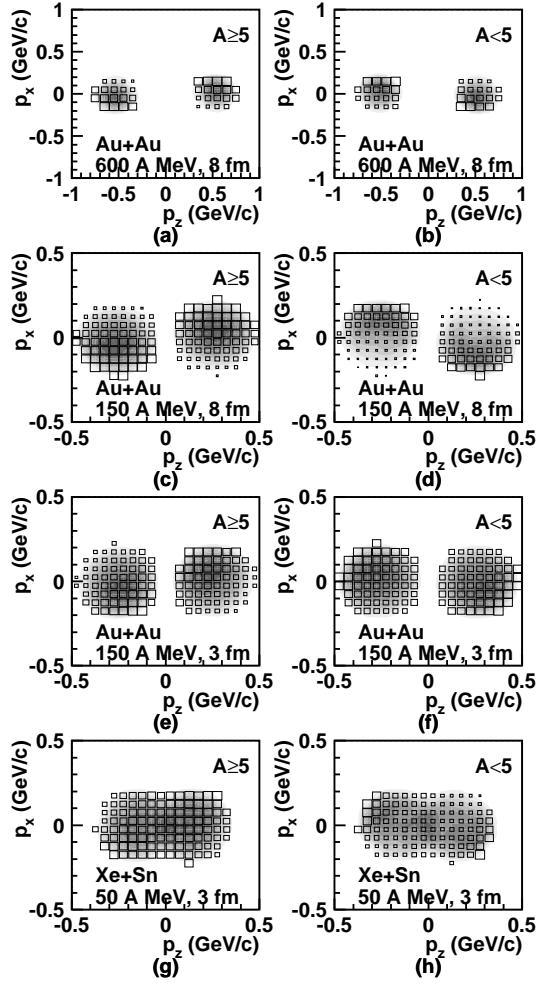}
$$
\caption{Momentum-space initial-final state correlations: initial
density of nucleons entrained finally in heavy fragments have been
plotted for the same reactions as in figure~\ref{xz_np.eps}.}
\label{pxpz_np}
\end{figure}

Usually, correlations between the initial momentum of the nucleons and 
their probability to end up in a given type of cluster are not considered 
as important. 
In general, phenomenological models assume that there are no
such correlations. However, the QMD calculations show quite strong
correlations. In figure~\ref{pxpz_np}, we present these correlations in
momentum space, in the same way as we did for the coordinate space in 
figure~\ref{xz_np.eps}.

At 600A~MeV we see that the heavy fragments are formed from nucleons
which have a momentum pointing away from the collision partner. 
This qualitative observation is independent of the impact parameter but the 
correlations are quantitatively stronger at smaller impact parameter
(cf. table~\ref{correl-moment-table}). At lower energy the correlation in 
$x$ direction is weakened and starting from E~=~150A~MeV supplemented by  
correlations in $z$ direction, especially important for the LFs.

While it is pretty easy to understand the origin of the transverse 
correlations at high energy, the longitudinal ones observed at
low energy are more subtle to grasp. They rely on two points:
\begin{itemize}
\item
One needs efficient collisions to produce light fragments
quickly (this can be seen at the best on figure~\ref{coll.eps}).
\item
At low beam energy, the Pauli blocking of the cross section is quite strong,
but less drastic for nucleons incoming with higher 
relative momentum. In other words, nucleons helped coherently by the Fermi 
momentum have a higher probability to scatter into an empty place of phase 
space.  
\end{itemize}
Combining these two points, one concludes that nucleons 
entering the reaction zone with large $|p_z|$ are more likely 
to end up in a light fragment than the others. At E~=~50A~MeV the Pauli 
blocking is severe: For almost all momenta one can reach in a nucleon-nucleon 
collision, the phase space cells are already occupied. Hence, the phase space 
opens up only for those nucleons incoming with particularly large values of 
$|p_z|$. 

To quantify the correlations in momentum space, we define the correlation 
numbers $C_{LF}^{p}$ and $C_{HF}^{p}$, exactly as we did in 
section~\ref{in-fin-correl-coord-space} for the correlations in 
coordinate space. 
In table~\ref{correl-moment-table}, we display some values of 
$C_{LF}^{p}$ and $C_{HF}^{p}$.

In general, the correlations in r and p-spaces are of the same order. 
In fact, r-space correlations are larger than p-space ones at high energies 
and smaller at low energies.

\subsection{Momentum-space correlations of projectile/target-like and
mid-rapidity fragments}
\label{in-fin-correl-moment-space-mrf-plf}
The correlation numbers in momentum space are presented in 
table~\ref{momentum-correl-part2}. 

As in coordinate space, strong correlations are observed, especially for 
the MRFs, which are essentially composed of participant nucleons 
(cf. section~\ref{in-fin-correl-coord-space-mrf-plf}). 
Consequently, we conclude that the participants do not really form a 
fireball (defined as an equilibrated system resulting from the destruction 
of {\em all} initial correlations).

\begin{figure}[t]
$$
\epsfbox{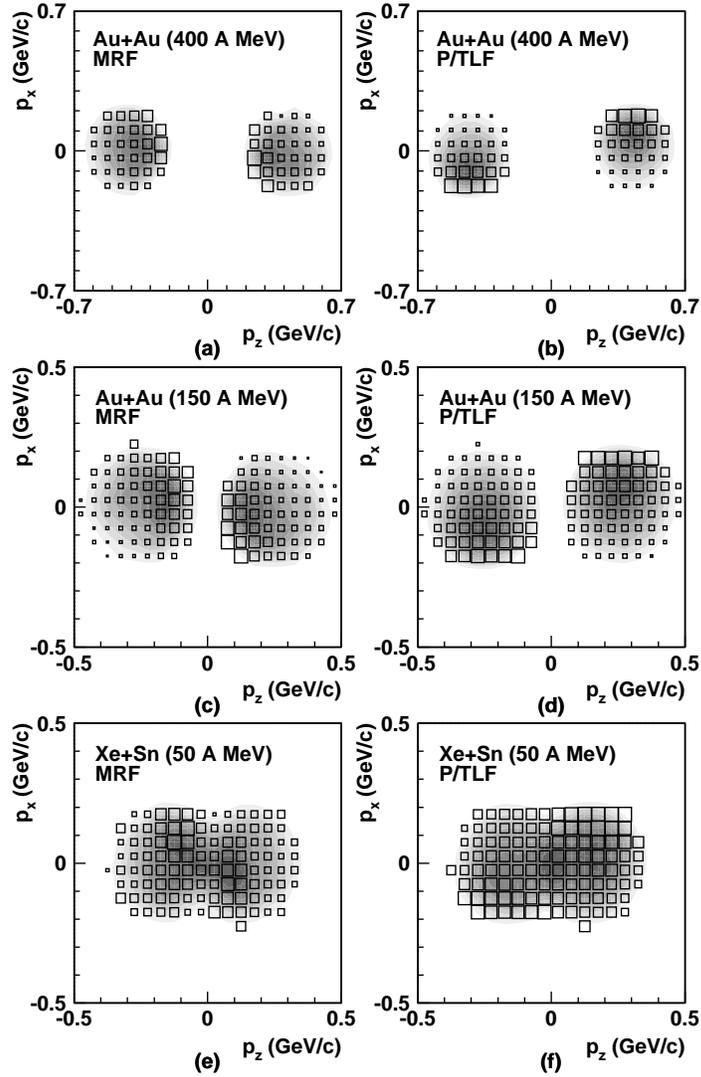}
$$
\caption{Momentum-space initial-final state correlations:
initial density of nucleons entrained finally in heavy fragments 
$(A\ge 5)$, separately for MRF and P/TLF. The 
same reactions as in figure~\ref{xz_lmf_2p_sb.ps} have been studied.}
\label{pxpz_lmf_2p_sb.ps}
\end{figure}

In figure~\ref{pxpz_lmf_2p_sb.ps}, we illustrate the momentum-space 
time-correlations both for MRFs and P/TLFs. P/TLFs mainly contain 
nucleons with an initial momentum pointing away from 
the collision partner. This type of correlation was already discovered
in section~\ref{in-fin-correl-moment-space}, in analyzing the HF's 
case\footnote{As P/TLF represent the largest part of the HF 
production they naturally exhibit the same correlations.}.

On the other hand, correlations of MRF's nucleons differ completely: 
Independently of the beam energy, MRFs are predominantly 
formed by nucleons which had initially a small energy in the nucleus-nucleus 
center of mass. As a matter of fact, nucleons with higher relative momentum 
possess a higher chance to perform a collision, due to their larger open phase 
space. Collisions provide a large momentum transfer and lower the
probability to find several fellow nucleons with about
the same final momentum, a necessary environment to form a fragment. 
Moreover, because of the relatively poor stopping at low energy, 
nucleons are only afforded a brief amount of time, during which they 
have to mix... if a MRF should be emitted. This constrain also favors 
nucleons with small relative energy, which have the possibility to stay
longer in contact.

Finally, we note that, as previously, we do not see any dramatic 
reduction or modification of the correlations when the model (B) is used 
instead of the model(A): From now on, we shall confidently restrict 
ourselves to the model (A).

\subsection{Catapult mechanism at low energy}
\label{catapult-section}
We shall now exploit the knowledge just gained about the longitudinal 
correlations in p-space at low energy to understand those in r-space. 

\begin{figure}[t]
$$
\epsfbox{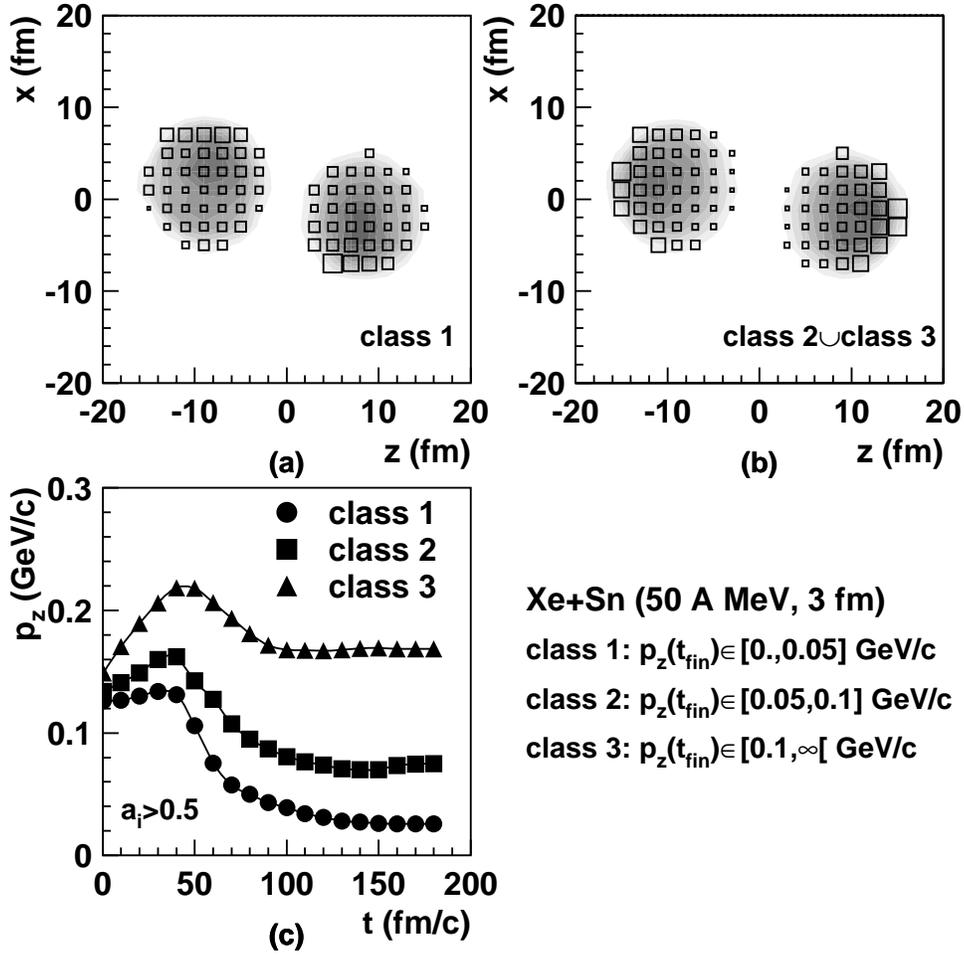}
$$
\caption{Xe~+~Sn reaction at E~=~50A~MeV, b~=~3~fm: initial position 
of those nucleons finally emitted in light fragments $(A\le 4)$
with (a) a small (b) large absolute longitudinal momentum per
nucleon. We also display (c) the time evolution of the
longitudinal momentum for three classes of nucleons.}
\label{katapult.ps}
\end{figure}

In fact, such longitudinal correlations are not completely new
to the expert: a while ago it has already been observed that the most 
energetic nucleons appear in the half-plane opposite to the impact 
parameter \cite{bondo}.   
This is surprising, because these nucleons have to cross the entire
nuclear system. To investigate the origin of this intriguing process 
we have selected, for the 50A~MeV reaction, nucleons which have 
finally the highest longitudinal momentum and searched for their initial 
position and momentum. Results are presented in figure~\ref{katapult.ps}. 

The particles with a moderate final momentum come from the 
participant zone, whereas the most energetic 
nucleons come from the back ends of the colliding nuclei,
in agreement with \cite{bondo}.
Initially these nucleons are not faster than their fellow nucleons,
but when they pass through the nucleus they get accelerated (cf panel(c)).
Only later, when they have to overcome the nuclear potential they loose 
momentum. 

We suggest the following reason: on the average, nucleons located 
at the back end surface of the projectile at initial time become 
the fastest ones some tens of fm/c later. One can then apply the arguments 
developed in section~\ref{in-fin-correl-moment-space}, where we have 
discussed in details why fast nucleons should emerge more easily as singles 
in low-energy reactions. 

Let us now clarify the physical mechanism: 
In a nucleus, nucleons are continuously moving and accelerated.
At a given time, nucleons located at the surface feel a density 
$\rho<\rho_{0}$ and are nearly at rest, as they have given 
their kinetic energy to climb the mean-field potential. Later on,
strongly re-accelerated towards the nucleus center by the force due to 
the density gradient, they become the fastest ones. Usually, they pass 
the center, arrive at the other end of the nucleus 
and are stopped again...If a 50A~MeV reaction takes place in between, the 
first stages of this scenario are basically unchanged:
at low beam energy, the density increases very little in the interaction 
region and remains almost constant until the nucleons have passed the entire 
reaction partner. At this time, they should normally be stopped. However, 
the relative momentum between them and the nucleons at the end of the 
reaction partner has now increased by the beam momentum and the density 
gradient decreased, due to dilution of nuclear-matter. As a result,
the force at the surface is no longer sufficient to stop those fast 
nucleons and they are just decelerated 
(see figure~\ref{katapult.ps}).

This catapult mechanism also explains the longitudinal correlations
observed in the MRF channel. Nucleons initially at the front end 
of the projectile are accelerated in the direction opposite to the
beam momentum. Very soon, they become the slowest ones in the system,
what increases their probability to emerge in MRFs, as demonstrated in
section~\ref{in-fin-correl-moment-space-mrf-plf}.

\subsection{Summary of the phase-space correlations}
\label{summar-correl}
Up to now we have investigated the existence of possible correlations 
between the initial position of nucleons in r and p-space and their final 
fate. Table~\ref{summary-correl} summarizes our present findings concerning 
the correlations.

It turns out that there are two rather distinct classes
of heavy fragments: those which are formed almost exclusively by 
projectile {\em or} target nucleons and those which are close to chemical 
equilibrium. For both classes we observe
strong initial-final state correlations, however different in their nature. 
At high energy, we recover the PS model and its associated transverse 
correlations in coordinate space with, nevertheless, strong 
supplementary correlations in momentum space. 

At lower energies, a semi-transparent regime was established: 
The PLFs (resp. TLFs) are mostly formed by nucleons which have traversed 
the entire target (resp. projectile). Due to the severe Pauli blocking of the collisions, 
projectile nucleons can indeed keep their correlations while 
traversing the target and vice versa. On the top of this mean behavior, 
longitudinal correlations in r and p-space were also discovered. For
instance, it appeared that MRF are formed preferably by nucleons having
a very small energy in the nucleus-nucleus center of mass.
In section~\ref{time-evol-of-reac}, we investigate further those reaction 
mechanisms.

Before this, we would like to interpret the 600A~MeV results
in view of the dependence of the transverse flow on the fragment mass.
Already the plastic ball collaboration has observed \cite{pb} 
that the directed transverse flow increases as a function of the mass number. 

In section~\ref{in-fin-correl-coord-space}, we have found that the
large fragments (even in central collisions) are made of spectator nucleons.
They do not pass a region of high density. Thus the directed transverse 
flow does not measure properties of the high density zone but the
gradient of the potential at its surface. Therefore the flow can
only measure the high density properties of nuclear matter which can be 
inferred from the gradient of the potential at the surface of the 
high density zone! 

In section~\ref{in-fin-correl-moment-space}, the situation was found to be
even worse. Indeed, the average {\it initial} transverse momentum of the 
fragment's nucleons differs from zero. This means that the above-mentioned 
transverse flow is partially generated during the collisions 
but {\em also} results from initial-final state correlations.
This questions another time the value of the observable flow as a messenger 
of the properties of the high density zone. 

These findings must
be confronted with the recent observation that the strange particle production
at that energy takes place in the high density region \cite{har}. 
Therefore these particles provide a more direct information about the high 
density zone.

\section{Time evolution of the reaction}
\label{time-evol-of-reac}
Now we have examined initial properties of the nucleons 
related to the nature of the produced fragments, we turn to the 
time-dependence of another set of observables, in order to understand 
better how those fragments emerge from the reaction.
A typical point we would like to clarify is how close the nucleons
finally forming a fragment have been in the past. Does a nucleus-nucleus
collision resemble a stone-stone collision? Or is it more an
evaporation-recondensation mechanism? Besides, we would like to understand
how a typical prefragment evolves under its interaction with the 
``external'' nucleons (defined as all the $A_P+A_T-A$ nucleons which do not 
belong to the same prefragment). 

\subsection{Radii and densities}
\label{radii-dens-section}
In relation~\ref{equation-cm}, we have defined the
center of mass ($\vec{R}$) and the mean momentum per nucleon ($\vec{P}$) 
of a given prefragment.
In the same spirit, we define for our purpose:  
\begin{itemize}
\item
$Rad(t)$, the radius of the prefragment as
\begin{equation}
Rad(t)=\sqrt{\frac{\sum_{i=1}^{A} (\vec{r}_{i}(t) -
        \vec{R}(t))^2}{A}} 
\end{equation}
\item
$Rad_{\perp}(t)$, the normalized transverse radius of the prefragment as
\begin{equation}
Rad_{\perp}(t)=\sqrt{\frac{3}{2}\frac{\sum_{i=1}^{A} 
        (\vec{r}_{i,\perp}(t) -
        \vec{R}_{\perp}(t))^2}{A}}\;,
\end{equation}
where the $\perp$ suffix indicates a transverse projection. 
The $3/2$ normalization factor allows a direct comparison with 
values of $Rad$.  
\item
$\rho_{int}(t)$, the density of the prefragment matter at 
its own center of mass as  
\begin{equation}
\rho_{int}(t)\ \propto\sum_{i=1}^{A} 
    e^{-\frac{(\vec{r}_{i}(t)-\vec{R}(t))^2}{2\rho_{0}^2}}
\end{equation}
\item 
$\rho_{ext}(t)$, the mean density of ``external'' matter at the
center of mass of the prefragment as
\begin{equation}
\rho_{ext}(t)\ \propto\sum_{i=A+1}^{A_{P}+A_{T}} 
    e^{-\frac{(\vec{r}_{i}(t)-\vec{R}(t))^2}{2\rho_{0}^2}}\;,
\end{equation}
where the sum runs on all external nucleons. In fact, the small value of 
$2\rho_{0}^2$ (2.17~$\mbox{fm}^2$) automatically favors external nucleons 
close to $\vec{R}$, without any supplementary condition.
\end{itemize}
One has of course 
\begin{equation}
\rho(\vec{R}(t),t)=\rho_{ext}(t)+\rho_{int}(t)
\end{equation}
but one will see that the distinction between internal and external
matter is quite fruitful. As usual, we average on all fragments of a given 
class and on a significant ensemble of QMD runs. 

\begin{figure}[t]
$$
\epsfbox{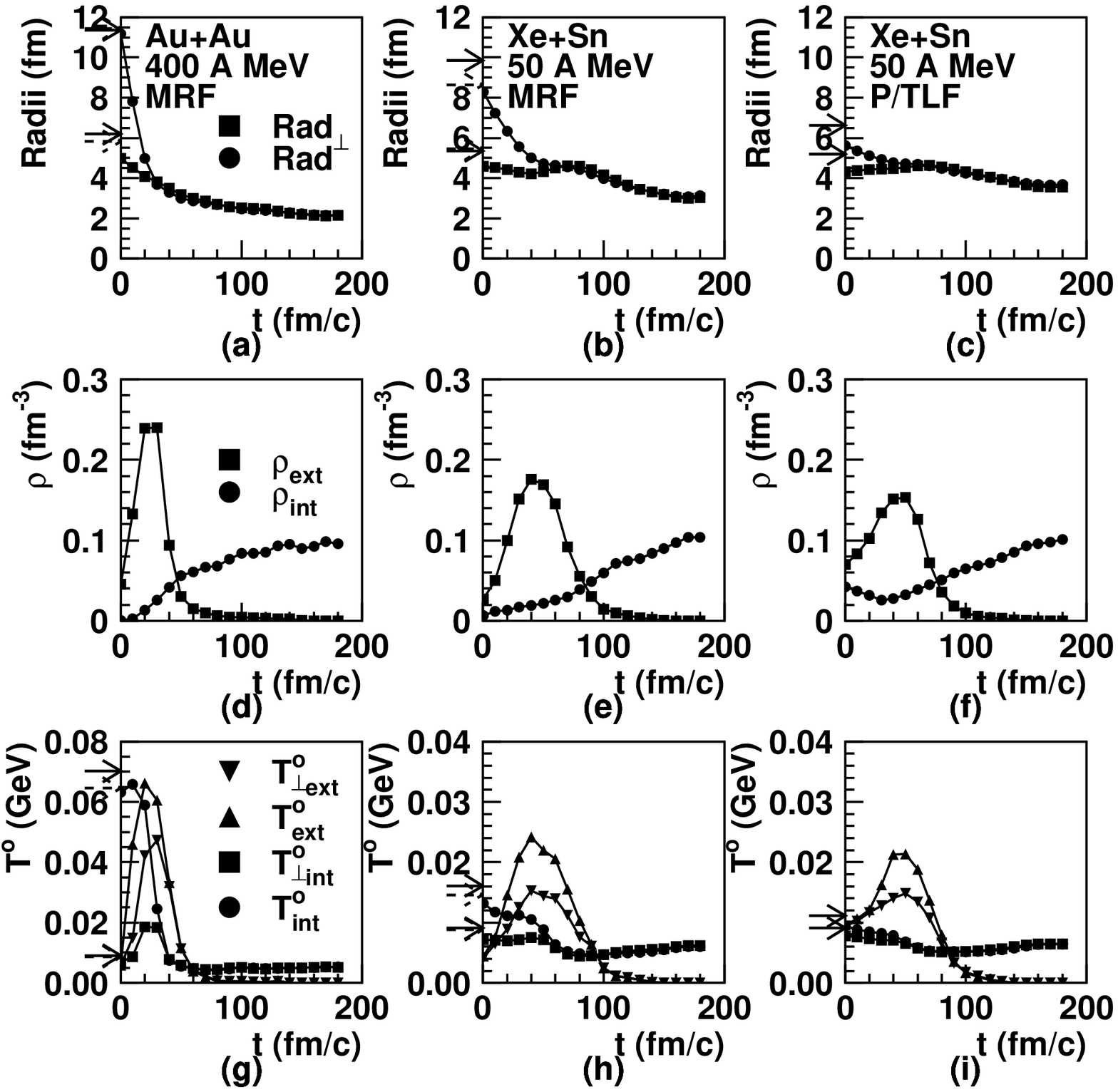}
$$
\caption{Time evolution of the radii (transverse and total), 
densities (external and internal) and temperatures (external and internal, 
transverse and total) are displayed at the top, middle and bottom 
respectively, for three classes of fragments selected in b~=~3~fm 
reactions: MRF in Au~+~Au, 400A~MeV (left), 
MRF and P/TLF in Xe~+~Sn, 50A~MeV (middle and right)---see
text for definitions. The plain and dashed arrows 
indicate the value of the equivalent quantity calculated within the 
one-body distribution assumption---see text for details.}
\label{orfrglmfb3_rad_rho_temp.eps}
\end{figure}

On figure~\ref{orfrglmfb3_rad_rho_temp.eps},
we plot the time evolution of these newly defined coordinate-space variables 
for b~=~3~fm reactions (radii at the top and densities in the middle). 
We address the case of MRF in Au~+~Au (400A~MeV), as well as 
MRF {\em and} PLF in Xe+Sn (50A~MeV), from left to right.

We mark with a {\em plain} arrow the initial value expected if no 
correlation were present, i.e. if fragments were made by taking nucleons 
randomly from projectile and target, but in the proper 
composition ($a_{i}$). Hence differences between  
plain arrows and actual values are due to the nonuniform distribution 
(over the target and projectile) of prefragment nucleons at initial time.
This is nothing but the initial-final state correlations already discussed.

For 400A~MeV there is no correlation in longitudinal direction but a 
small one in transverse direction. MRFs stem to a large extent from the 
participant region which has a smaller transverse size than the combined 
system of participants and spectators. If we go down to the lower energy 
we observe initial correlations also in longitudinal direction as already 
seen in figure~\ref{xz_np.eps}\ and \ref{xz_lmf_2p_sb.ps}. 

The {\em dashed} arrow marks the initial value expected if one takes 
the nucleons (in the proper composition ($a_{i}$)) randomly from a 
one-body distribution\footnote{Examples of such one-body distributions
have been given in figure~\ref{xz_lmf_2p_sb.ps}.} made of all nucleons 
{\em belonging to the same class of prefragments} (MRF or PLF). 
Any difference between the dashed arrow and the actual value gives
a new insight: it indicates the presence of true many-body correlations, 
for example that fragment nucleons have been already close (in coordinate 
space) initially. 

Introducing this refinement, we see that part of the discrepancies
between the pure statistical model (plain arrows) and the actual values 
can be understood in terms of one-body effects (dashed arrows), but 
only part of it, pleading for some many-body correlations.  

Let us examine the time-evolution: at 400A~MeV, it takes about 
30~fm/c---50~fm/c at 50A~MeV---until the transverse and the longitudinal 
sizes of the prefragment match. In fact, the radii decrease while 
nucleons from different regions of projectile and target join to form 
a fragment (the higher the energy the smaller is the probability for 
the prefragment nucleons to have been already close together initially).
One can conclude that those coordinate-space 
correlations are one but not the essential reason for fragment formation.

This is even better reflected by the increase of the 
prefragment internal density: at small times $\rho_{int}$ is still quite 
low and strongly dominated by $\rho_{ext}$, even at the center of
mass of the prefragment itself. Let us also notice that for the 400A~MeV 
reaction, $\rho_{ext}$ exceeds by far the normal nuclear-matter density.

After 100~fm/c (for the 400A~MeV reaction) or 150~fm/c 
(for the 50A~MeV reaction), no other nucleon can be found around
the prefragment's members. At that time the prefragments have
separated from each other and can be considered as real and dense fragments,
easily identified by a minimum spanning tree algorithm.

From the reactions we have studied in this section, one can conclude that 
nucleus-nucleus and stone-stone collisions have little in common,
at least as far as the mechanisms for fragment formation are concerned.
In the next section, we demonstrate that an evaporation-recondensation 
scenario is irrelevant as well.

\subsection {$P_{z}$ and effective temperatures}
\label{pz-temp-section}
In a dedicated study, it appeared that at low energy (50A~MeV),
$\overline{P_{z}}$, does practically not change in the course of time.
At higher energy, this statement has to be attenuated, but is still
relevant at 400A~MeV... Is it really so that the average prefragment 
passes through the high density/temperature zone without being affected 
by it? To obtain new insights into this question, a natural way
consists in comparing their {\em respective} temperatures (or
excitation energies). For this purpose, we introduce:

\begin{itemize}
\item
$T_{int}(t)$, the effective temperature of the prefragment as
\begin{equation}
T_{int}(t)=\frac{\sum_{i=1}^{A} 
(\vec{p}_{i}(t)-\vec{P}(t))^2}{3.m_{N}.A} 
\end{equation}
where $m_{N}$ is the nucleon mass. Note that $T_{int}(t)$ is an effective 
temperature of the fragment which agrees with the true temperature only
in the case of a Maxwellian momentum-distribution. However, being the
second moment of the momentum distribution it gives, in any case, a measure 
for the excitation energy/nucleon.

\item 
$T_{\perp~int}(t)$, the effective transverse temperature of the prefragment, 
defined as above by taking only into account the transverse degrees of 
freedom.

\item
$\vec{P}_{ext}(t)$, the mean momentum of the external matter as
\begin{equation}
\vec{P}_{ext}(t)=\frac{\sum_{i=1}^{A_{ext}(t)} \vec{p}_{i}(t)}
{A_{ext}(t)}\;, 
\end{equation}
where $A_{ext}(t)$ is the number of external nucleons {\em located} 
within 3~fm from the prefragment center of mass 
$\vec{R}(t)$.

\item
$T_{ext}(t)$, the effective temperature of the external matter as
\begin{equation}
T_{ext}(t)=\frac{\sum_{i=1}^{A_{ext}(t)} 
(\vec{p}_{i}(t) -\vec{P}_{ext}(t))^2 }{3.m_{N}.A_{ext}(t)} 
\end{equation}
where the sum over i is performed on the same nucleons as for 
$\vec{P}_{ext}(t)$. Again, this quan\-ti\-ty is to be considered as a 
probe of the medium excitation-energy.

\item
$T_{\perp~ext}(t)$, the effective transverse temperature of the external
matter, i.e. the second moment of the momentum distribution in
transverse direction.
\end{itemize}

As usual, the double averaging is performed. At the bottom of 
figure~\ref{orfrglmfb3_rad_rho_temp.eps}, these various temperatures 
are displayed for the same reactions as
those considered in section~\ref{radii-dens-section}

The external temperatures show the expected behavior: When projectile and 
target nucleons interact, the system heats up and the variance of the
momentum distribution increases. In an equilibrated system, the transverse 
and total temperature should be equal. This is definitively not the case in 
our results, until 40~fm/c for the 400A~MeV reaction and until 90~fm/c
for the 50A~MeV reaction. This lack of external equilibrium is
consistent with all the correlations found so far. 
Finally the temperature lowers because the system disassembles 
and---depending of their velocity---the external nucleons leave the 3~fm 
sphere around the moving center of mass of the prefragment. Now, only 
prefragment nucleons are present in this sphere and therefore the 
external variances become identically zero. 

The internal temperatures show a completely different behavior:
Initially the prefragment nucleons have a lower variance than 
expected from a random selection of those nucleons. As expected, 
MRF nucleons have a quite large variance in the beam direction. In the
course of the time, the internal temperatures decrease even before the 
system expands. This behavior completely contradicts our intuition as 
well as our experience from thermodynamics. In a thermodynamical language: 
the fragments cool down by giving energy to the already hot environment.
It is obvious that these findings forbid to consider the second
moment as a common temperature, in the sense of Gibbs' 
criteria: $T_{int}=T_{ext}$.  

To understand what is actually happening, one uses---once again---a
feedback argument: In the end, only those nucleons with small relative 
momenta form clusters. In the course of the reaction, it happens that 
some nucleons are scattered into the vicinity (in phase space) of one nucleon 
or a cluster of nucleons. With respect to these nucleons the variance of the 
relative momentum is smaller after the collision than before. If the nucleons 
do not scatter once more they may leave the interaction region as a group and 
may form a cluster. If a nucleon scatters once more it will not be 
entrained in this cluster and hence will not affect the evaluation of 
$T_{int}$.

Thus one may conclude that:
\begin{itemize}
\item
Initial correlations are present: Geometry and relative momenta
determine how probably nucleons are scattered out of or into
certain momentum-space regions. 

\item
Although located at the same place in coordinate space as
the surrounding nucleons, the prefragment nucleons have a much lower 
variance of their relative momenta. The fragment formation is hence
a fluctuation on the level of the one-body phase space distribution.

\item 
Fragments are formed 
(a) by nucleons which are already close in phase space at the beginning of 
the reaction and keep this correlation during the reaction
and (b) by nucleons which are brought by accident into the same phase 
space region (this may be a consequence of two-body scattering or of the 
potential interaction for MRF). In other words,  
fragmentation is a mixture of conserving initial correlations
(for projectile and target-like fragments) and of building up new 
correlations in the course of the interaction (for mid-rapidity fragments).
\end{itemize} 

\section{Conclusions}
\label{conclusions}
Using the QMD model, we have extensively studied the process of 
multifragmentation in heavy ion induced reactions. We have validated 
the QMD approach by showing that the
presently available data are in even better agreement
with the calculation than expected in view of approximations made in
the model, especially at low energy. Also, we have checked and 
explained in details that the qualitative conclusions given hereafter 
depend very little on the symmetrized character of the microscopic NN 
interaction.   

At all energies, we have found that correlations in coordinate and 
momentum space (the memory of the entrance channel) are preserved during the 
time-evolution of the reaction. In other words, the probability for
nucleons to be finally entrained in fragments depends on their initial
position in phase space. However, the regions of the initial phase space
which lead to a high probability of ending in a fragment are quite different
for different energies and different impact parameters. At the
highest energy investigated we recover the well-established participant
spectator model, supplemented by strong correlations
in momentum space. At the lowest energy we find a situation 
reminiscent of deep inelastic collisions for large impact parameters and
a semi-transparent regime for nearly central reactions. In between, there 
is a wide range of time-correlations which have been explained in physical 
terms.

According to our analysis, fragments are formed by fluctuations in phase 
space: initially {\em or} during the time evolution of the 
reaction some nucleons are {\em or} come by chance into the same phase 
space region. If these nucleons do not suffer a hard scattering at further 
time, they have a large  probability to emerge from the reaction zone as a 
fragment. In the first case, a fragment dominantly formed from either 
projectile or target nucleons will be emitted, in the second, one
will observe a fragment containing mostly identical admixtures of
projectile and target nucleons.

These two mechanisms coexist {\em whatever} the energy. In general, 
however, those fragments composed mainly of projectile {\em or} target
nucleons dominate. As the energy is decreased, the number of fragments 
which have about the same number of projectile and target nucleons 
increases but remains in the minority. 

During the reactions, the internal "temperature" (the second moment 
of the momentum distribution of prefragment nucleons) is much lower than 
the external ``temperature''. This proves that fragment formation
is really due to fluctuations, that is quantities which are not calculable 
in a one-body theory. This may also explain why in thermal models 
the "temperature" determined from the isotopic yield is always lower 
than the one determined from the spectra.

Apart maybe at very high energy and very small impact parameter, we have 
found no evidence that the system or part of the system comes close to 
equilibrium, nor do we find evidence that clusters are formed 
at densities expected for the spinodal transition... In fact, the closest 
subsystem to equilibrium is the spectator matter in peripheral reactions 
at high energies. For this case, addressed in a specific work~\cite{puri}, 
several observables follow the trend one expects from thermodynamics. The 
key variable, however, the width of the momentum-space distribution as a 
function of the fragment mass, undoubtedly departs from this trend, 
in agreement with experiment, and is too large to be consistent with a 
possible excitation energy of the system. 
  
In this paper, we have concentrated on symmetric systems. At low
energies we do not expect that asymmetric systems behave very differently. 
At higher energies, where the geometry cuts play an important role, there 
may be differences, i.e. in a asymmetric system the projectile may drill
a hole through the target nucleus. This would have of course consequences
for the geometrical structure of the spectator part and may be investigated 
in an upcoming publication.
    
\section*{Acknowledgments}
This work was partially supported by the National Science Fundation under 
grant PHY-9403666.

\clearpage 
\begin{table} [th]
\caption{$\overline{E}_{rat}$ vs. $b$ for various systems}
\begin{center}
\begin{tabular} {|p{2.0cm}|p{2.1cm}|p{2.1cm}|p{2.1cm}|p{2.1cm}|}
\hline
b$\backslash$Reaction 
& Xe+Sn(50)& Au+Au(50)& Au+Au(150)& Au+Au(400)\\
\hline
 0 fm & 0.75 & 0.83 & 1.45 & 1.67 \\
 3 fm & 0.66 & 0.71 & 1.02 & 1.11 \\
 6 fm & 0.47 &      &      &      \\
 8 fm &      & 0.37 & 0.25 & 0.26 \\
\hline
\end{tabular}
\label{table-erat}
\end{center}
\end{table}

\begin{table} [th]
\caption{Fraction of MRFs among HFs, evaluated in both models: 
model (A)$|$ model(B)}
\begin{center}
\begin{tabular} {|p{1.5cm}|p{2.1cm}|p{2.1cm}|p{2.1cm}|p{2.1cm}|}
\hline
b$\backslash$Reac.& 
Xe+Sn(50) & Au+Au(50)  & Au+Au(150) & Au+Au(400)\\ 
\hline
 0 fm & $0.28|0.41$ & $0.32|0.45$ & $0.42|0.64$ & $0.55|0.73$ \\ 
 3 fm & $0.25|0.35$ & $0.29|0.41$ & $0.32|0.49$ & $0.27|0.43$ \\ 
 6 fm & $0.16|0.23$ &             &             &             \\ 
 8 fm &             & $0.17|0.25$ & $0.13|0.19$ & $0.05|0.07$ \\ 
\hline
\end{tabular}
\label{chemical-equil}
\end{center}
\end{table}

\begin{table} [th]
\caption{Qualitative dependence of the chemical equilibration 
on the physical parameters}
\begin{center}
\begin{tabular} {|p{2cm}|p{2cm}|p{2cm}|}
\hline
dependence&variable&fixed\\
\hline
sharp &  b                       & E$>$   \\
smooth&  b                       & E$<$   \\
smooth&  E                       & b=0    \\
smooth&  $A_{T}+A_{P}$           & E$<$,b \\
linear&  $\bar{\theta}_{deflect}$ & E,b    \\
\hline
\end{tabular}
\label{table-dep-dable}
\end{center}
\end{table}

\begin{table} [th]
\caption{$\overline{\hat{\rho}_{LF}}$ and $\overline{\hat{\rho}_{HF}}$, in \%} 
\begin{center}
\begin{tabular} {|p{3.6cm}|p{1.6cm}|p{1.6cm}|p{1.6cm}|p{1.6cm}|}
\hline
Reaction &\multicolumn{2}{c|} {$b_<$}&
\multicolumn{2}{c|} {$b_>$}\\ \cline{2-5}
& LF & HF & LF & HF \\ 
\hline
Au+Au (600A~MeV) & 93 & 07 & 55 & 45 \\
Au+Au (400A~MeV) & 88 & 12 & 50 & 50 \\
Au+Au (150A~MeV) & 63 & 37 & 36 & 64 \\
Xe+Sn (50A~MeV)  & 33 & 67 & 24 & 76 \\
\hline
\end{tabular}
\label{table-meanrho}
\end{center}
\end{table}

\begin{table} [th]
\caption{$C_{LF}^{r}$ and $C_{HF}^{r}$}
\begin{center}
\begin{tabular} {|p{3.6cm}|p{1.6cm}|p{1.6cm}|p{1.6cm}|p{1.6cm}|}
\hline
Reaction &\multicolumn{2}{c|} {$b_<$}&
\multicolumn{2}{c|} {$b_>$}\\ \cline{2-5}
& LF & HF & LF & HF \\ 
\hline
Au+Au (600A~MeV) & 0.06 & 0.75 & 0.48 & 0.54 \\
Au+Au (400A~MeV) & 0.05 & 0.41 & 0.45 & 0.43 \\
Au+Au (150A~MeV) & 0.06 & 0.09 & 0.30 & 0.16 \\
Xe+Sn (50A~MeV)  & 0.29 & 0.13 & 0.26 & 0.08 \\
\hline
\end{tabular}
\label{table-correl-spac}
\end{center}
\end{table}

\begin{table} [th]
\caption{$\overline{\hat{\rho}_{MRF}}$ and 
$\overline{\hat{\rho}_{P\cup LF}}$, in \% and for both models: 
$\mbox{model (A)}|\mbox{model (B)}$}
\begin{center}
\begin{tabular} {|p{3.6cm}|p{1.6cm}|p{1.6cm}|p{1.6cm}|p{1.6cm}|}
\hline
Reaction &\multicolumn{2}{c|} {$b_<$}&
\multicolumn{2}{c|} {$b_>$}\\ \cline{2-5}
& MRF & P$\cup$TLF  & MRF & P$\cup$TLF \\ 
\hline 
Au+Au (400A~MeV) &  $02|05$ & $09|07$ & $0.2|0.4$ & $51|50$ \\
Au+Au (150A~MeV) &  $11|17$ & $29|22$ & $01|02$   & $64|62$ \\
Xe+Sn (50A~MeV)  &  $14|19$ & $56|51$ & $07|10$   & $69|67$ \\
\hline
\end{tabular}
\label{mean-rho-part2}
\end{center}
\end{table}

\begin{table} [th]
\caption{$C^{r}_{MRF}$ and $C^{r}_{P\cup TLF}$ for both models: 
$\mbox{model (A)}|\mbox{model (B)}$}
\begin{center}
\begin{tabular} {|p{3.6cm}|p{1.6cm}|p{1.6cm}|p{1.6cm}|p{1.6cm}|}
\hline
Reaction &\multicolumn{2}{c|} {$b_<$}&
\multicolumn{2}{c|} {$b_>$}\\ \cline{2-5}
& MRF & P$\cup$TLF  & MRF & P$\cup$TLF \\ 
\hline
Au+Au (400A~MeV) & $0.33|0.23$ & $0.55|0.54$ & $1.07|0.76$ & $0.43|0.39$ \\
Au+Au (150A~MeV) & $0.28|0.13$ & $0.15|0.17$ & $0.69|0.35$ & $0.17|0.14$ \\
Xe+Sn (50A~MeV)  & $0.59|0.43$ & $0.11|0.11$ & $0.75|0.64$ & $0.10|0.10$ \\
\hline
\end{tabular}
\label{space-correl-part2}
\end{center}
\end{table}

\begin{table} [th]
\caption{$C_{LF}^{p}$ and $C_{HF}^{p}$}
\begin{center}
\begin{tabular} {|p{3.6cm}|p{1.6cm}|p{1.6cm}|p{1.6cm}|p{1.6cm}|}
\hline
Reaction &\multicolumn{2}{c|} {$b_<$}&
\multicolumn{2}{c|} {$b_>$}\\ \cline{2-5}
& LF & HF & LF & HF \\ 
\hline
Au+Au (600A~MeV) & 0.03 & 0.44 & 0.28 & 0.31 \\
Au+Au (400A~MeV) & 0.05 & 0.36 & 0.36 & 0.34 \\
Au+Au (150A~MeV) & 0.16 & 0.26 & 0.58 & 0.31 \\
Xe+Sn (50A~MeV) & 0.3 & 0.13 & 0.47 & 0.15 \\
\hline
\end{tabular}
\label{correl-moment-table}
\end{center}
\end{table}

\begin{table} [th]
\caption{$C_{MRF}^{p}$ and $C_{P\cup TLF}^{p}$ for both models: 
$\mbox{model (A)}|\mbox{model (B)}$}
\begin{center}
\begin{tabular} {|p{3.6cm}|p{1.6cm}|p{1.6cm}|p{1.6cm}|p{1.6cm}|}
\hline
Reaction &\multicolumn{2}{c|} {$b_<$}&
\multicolumn{2}{c|} {$b_>$}\\ \cline{2-5}
& MRF & P$\cup$TLF  & MRF & P$\cup$TLF \\ 
\hline
Au+Au (400A~MeV) & $0.34|0.22$ & $0.51|0.46$ & $0.76|0.65$  & $0.34|0.32$ \\
Au+Au (150A~MeV) & $0.48|0.26$ & $0.37|0.35$ & $1.02|0.71$  & $0.33|0.30$ \\
Xe+Sn (50A~MeV)  & $0.27|0.22$ & $0.15|0.15$ & $0.30|0.26$  & $0.16|0.15$ \\
\hline
\end{tabular}
\label{momentum-correl-part2}
\end{center}
\end{table}

\begin{table} [th]
\caption{Phase-space correlations: The first and second line at each 
entry respectively summarize our findings for large and small impact 
parameter reactions. $++$ stands for strong correlation along 
the axis considered, $--$ for strong anticorrelation and the arrows point 
into the direction of decreasing energy.}
\begin{center}
\begin{tabular} {|p{4.0cm}|p{1.5cm}|p{1.5cm}|p{1.5cm}|p{1.5cm}|}
\hline
Fragment $\backslash$  Correl. & $R_{\perp}$ & $R_{\parallel}$  & $P_{\perp}$ & 
$P_{\parallel}$ \\
\hline
LF & $++\rightarrow-$ & $0\rightarrow-$ & $++\rightarrow+$ & $0\rightarrow--$ \\
~ & $0\rightarrow0$ & $0\rightarrow--$ & $0\rightarrow0$ & $0\rightarrow--$ \\
\hline
PLF/TLF & $--\rightarrow0$ & $0\rightarrow0$ & $--\rightarrow-$ & $0\rightarrow0$ 
\\
~ & $--\rightarrow0$ & $0\rightarrow0$ & $--\rightarrow-$ & $0\rightarrow0$ \\
\hline
MRF & $++\rightarrow0$ & $0\rightarrow++$ & $++\rightarrow+$ & 
$+\rightarrow+$ \\
~ & $+\rightarrow0$ & $0\rightarrow++$ & $0\rightarrow0$ & $++\rightarrow+$ \\
\hline\end{tabular}
\label{summary-correl}
\end{center}
\end{table}

\clearpage 
\epsfxsize 4.5in

\clearpage 
\epsfxsize 5.5in

\clearpage 
\epsfxsize 5in

\clearpage
\epsfxsize 5.5in

\clearpage 
\epsfxsize 5.5in

\clearpage
\epsfxsize 4.5in

\clearpage 
\epsfxsize 3.in

\clearpage
\epsfxsize 5in

\clearpage 
\epsfxsize 5.5in

\clearpage 
\epsfxsize 5in

\clearpage 
\epsfxsize 5.5in

\clearpage 
\epsfxsize 5.5in

\clearpage 
\epsfxsize 5.5in

\clearpage 
\epsfxsize 5.5in

\clearpage 
\epsfxsize 5.5in

\clearpage 
\epsfxsize 4.in

\clearpage 
\epsfxsize 5.5in

\clearpage 
\epsfxsize 4.in

\clearpage 
\epsfxsize 5.5in

\clearpage 
\epsfxsize 5.5in

\begin{thebibliography}{99}

\bibitem{fried}G. Friedl\"ander et al., Phys. Rev.94, 727 (1954)

\bibitem{WW} A. I. Warwick {\it et al.}, Phys. Rev. C {\bf 27}, 1083 (1983).  

\bibitem{hi} A. S. Hirsch {\it et al.}, Phys. Rev. C {\bf 29}, 508 (1984).

\bibitem{fi} M.E. Fischer, Physica {\bf 3} 255 (1967) and Rep. Prog. Phys.
{\bf 67}, 255  (1967)

\bibitem{mor} L. G. Moretto and G. J. Wozniak, Annual Reviews of
Nuclear and Particle Science, J. D. Jackson, Ed., {\bf 43}, 379 (1993).  

\bibitem{mi} U. Milkau, GSI Report 91-34 (1991).  

\bibitem{WT} W. Trautmann, Proceedings of the International Symposium
``Towards a Unified Picture of Nuclear Dynamics", Nikko, Japan, 1991.  


\bibitem{AH} J. Aichelin, J. H\"ufner and R. Ibarra, Phys. Rev. C {\bf 30}, 
107 (1984).  

\bibitem{je} 
S. C. Jeong {\it et al.}, Phys. Rev. Lett. {\bf 72}, 3468 (1994);
U. Sodan, Thesis, U. of Heidelberg, 1994 (unpublished).

\bibitem{gro} D. H. E. Gross, Rep. Prog. Phys. {\bf 53}, 605 (1990), 
and references therein.
 
\bibitem{bon} J. B. Bondorf, R. Donangelo, I. N. Mishustin, C. J. Pethick,
H. Schulz, K. Sneppen, Nucl. Phys. {\bf A443}, 321 (1985).
 
\bibitem{sto} D. Hahn and H. St\"ocker Nucl. Phys. {\bf A476}, 717 (1988).
  

\bibitem{bot} A. S. Botvina {\it et al.}, GSI 94-36 (1994).  


\bibitem{hs} W. C. Hsi {\it et al.}, preprint MSUNSCL - 930, 1994; 
             M. Begemann Blaich and W. Trautmann, GSI Nachrichten 05 - 94. 
 
\bibitem{aic} 
J. Aichelin, Phys. Rep. {\bf 202}, 233 (1991), and references therein. 

\bibitem{aich} 
J. Aichelin, Prog. Nucl Part. Phys. {\bf 30}, 191 (1993)

\bibitem{har}
C. Hartnack et al. Nucl Phys. {\bf A580}, 643 (1994)

\bibitem{bet}         G.F. Peaslee et al., Phys. Rev. C49 (1994) R2271

\bibitem{ps}      
G. Westfall, J. Gosset, P.J. Johanson, A.M. Poskanzer                   
W.G. Meyer, H.H Gutbrod, A. Sandoval and R. Stock, Phys. Rev. Lett.            
{\bf 37} (1976) 1202.

\bibitem{MSU} M.B. Tsang, private communication and proceedings
of the international workshop XXII on gross properties of nuclei
and nuclear excitations, Hirschegg, Austria, 1994, ed. by H. Feldmeier
and W. N\"orenberg

\bibitem{sang} Sangster, private communication, Phys. Rev. {\bf C51}, 1280
(1995)

\bibitem{bb} M. Begemann Blaich et al., Phys. Rev. {\bf C 48}, 610 (1993)

\bibitem{bbl} M. Begemann Blaich et al., Phys. Lett. {\bf B 298}, 27 (1993)

\bibitem{fm} H. Feldmeier Nucl. Phys. {\bf A586}, 493 (1995)

\bibitem{ho} A. Ono et al., Phys. Rev. Lett {\bf 68} 2898 (1992)

\bibitem{fmm} H. Feldmeier, private communication

\bibitem{aicph} J. Aichelin, Phys. Rev. {\bf C33 (1986) 537}

\bibitem{pb}  K.G.R. Doss et al., Phys. Rev. Lett. {\bf 59} (1987) 2720

\bibitem{in} 
B. Borderie {\it et al.}, INDRA collaboration, GANIL, private 
communication, and to be published.

\bibitem{eos}
Preliminary data from the EOS detector also support this conclusion.  

\bibitem{bondo}  J.P. Bondorf et al. Nucl. Phys. {\bf A 333} (1980) 285

\bibitem{brad} F.P. Brady et al., Phys. Rev. {C50} (1994) R525

\bibitem{dreu} J. Dreute et al., Phys. Rev. {\bf C44} (1991) 1057 

\bibitem{ppp} W. Trautmann et al., private communication

\bibitem{puri} R.P. Puri, P.-B. Gossiaux, Ch. Hartnack, and J. Aichelin, 
        accepted for publication in Nucl. Phys. {\bf A}
\end{thebibliography}
\end{document}